 \documentclass[final,5p,twocolumn,authoryear]{elsarticle}

\usepackage{upgreek}
\usepackage{amssymb}
\usepackage{amsthm}
\usepackage{amsmath,amsfonts,thmtools}
\usepackage{subfigure}
\usepackage{xcolor}
\usepackage{stmaryrd}
\usepackage{algorithmic}
\usepackage{algorithm}
\usepackage{ulem}
\usepackage{url}

\newcommand{\revise}[1]{{\color{black} #1}}

\renewcommand{\em}{\it}
\def\fee{\upphi}
\newcommand{\RU}{R}
\newcommand{\Ybar}{\bar{X}}

\newcommand{\K}{{\sf K}}
\newcommand{\etabar}{{\eta}}
\newcommand{\newP}[1]{\noindent{\bf #1:}}
\newcommand{\Sbar}{\bar{S}}
\newcommand{\ud}{\mathrm{d}}
\def\Re{\mathbb{R}}

\def\argmin{\mathop{\text{\rm arg\,min}}}

\def\Sec#1{Sec.~\ref{#1}}
\def\Fig#1{Fig.~\ref{#1}}

\def\eqdef{\mathrel{:=}}

\def\transpose{{\hbox{\rm\tiny T}}}

\def\clZ{{\cal Z}}

\newcommand{\backward}[1]{\overset{\shortleftarrow}{#1}}
\newcommand{\Ybarbar}{\bar{Y}}
\newcommand{\SN}{S^{(N)}}
\def\Sec#1{Sec.~\ref{#1}}
\def\Thm#1{Thm.~\ref{#1}}
\def\Prop#1{Prop.~\ref{#1}}

\newcommand{\phiepsN}{{\phi}^{(N)}_\epsilon}
\newcommand{\phieps}{{\phi}_{\epsilon}}

\newcommand{\rhoepsN}{\rho_\epsilon^{(N)}}
\newcommand{\rhoeps}{\rho^{(\epsilon)}}
\newcommand{\Teps}{{T}_{\epsilon}}

\newcommand{\TepsN}{{{T}^{(N)}_\epsilon}}

\newcommand{\pr}{\rho}
\newcommand{\neps}{{n}_{\epsilon}}

\newcommand{\mN}{m^{(N)}}
\newcommand{\SigN}{\Sigma^{(N)}}

\newcounter{rmnum}

\newcounter{anum}

\newcommand{\Ten}{{\sf T}}
\newcommand{\preps}{\pr_\epsilon}

\newcommand{\hvec}{{\sf h}}
\newcommand{\phivec}{{\sf \Phi}}

\newcommand{\Xbar}{\bar{X}}

\newcommand{\NN}{\mathcal{N}}
\def\Kbar{\bar{\sf K}}
\newcommand{\mbar}{\bar{m}}

\newcommand{\Sigmabar}{\bar{\Sigma}}

\newcommand{\Ricc}{\text{Ricc}}

\def\FRAC#1#2#3{\genfrac{}{}{}{#1}{#2}{#3}}
\def\half{{\mathchoice{\FRAC{1}{1}{2}}%
		{\FRAC{2}{1}{2}}%
		{\FRAC{3}{1}{2}}%
		{\FRAC{4}{1}{2}}}}
	
\newcommand{\calN}{\mathcal{N}}

\newcommand{\calF}{\mathcal{F}}

\newcommand{\Zdim}{m}
\newcommand{\Bdim}{q}

\newcommand{\Bbar}{\bar{B}}
\newcommand{\Wbar}{\bar{W}}

\declaretheorem[numberwithin=section]{theorem}

\declaretheorem[sibling=theorem]{proposition}

\declaretheorem[sibling=theorem]{remark}

\declaretheorem[sibling=theorem]{example}
\declaretheorem{assumption}

\journal{Annual Reviews in Control}
\begin{document}




\title{A Survey of Feedback Particle Filter and related \\ Controlled Interacting
  Particle Systems (CIPS)\tnoteref{t1}}

\tnotetext[t1]{The authors gratefully acknowledge the continued
  support of the National Science Foundation through the current grant
  1761622, and the past grants 1435555 and 1334987.}

\tnotetext[t1]{The authors' prior research---reported here---is based on contributions
  from a number of graduate students and collaborators at the University of
  Illinois at Urbana-Champaign.  These collaborations are gratefully acknowledged.}


\author[1]{Amirhossein Taghvaei}
\ead{amirtag@uw.edu}
\author[2]{Prashant G. Mehta\corref{cor}}
\ead{mehtapg@illinois.edu}
\cortext[cor]{Corresponding author.}

\affiliation[1]{organization={William E. Boeing Department of Aeronautics \& Astronautics, University of Washington},
            city={Seattle},
            postcode={98195}, 
            state={WA},
            country={USA}}

\affiliation[2]{organization={Coordinated Science Laboratory and the 
    Department of Mechanical Science and Engineering, University of Illinois},
	city={Urbana-Champaign},
	postcode={61801}, 
	state={IL},
	country={USA}}

\begin{abstract}
In this survey, we describe controlled interacting particle systems
(CIPS) to approximate the solution of the optimal filtering and the
optimal control problems.  Part I of the survey is focussed on the
feedback particle filter (FPF) algorithm, its derivation based on optimal transportation theory, and
its relationship to the ensemble Kalman filter (EnKF) and the conventional 
sequential
	importance sampling-resampling (SIR) particle
filters.  The central numerical problem of FPF---to approximate the solution of
the Poisson equation---is described together with the main solution
approaches.  An analytical and numerical comparison with the SIR particle filter is
given to illustrate the advantages of the CIPS approach.  Part II
of the survey is focussed on adapting these algorithms for the problem
of reinforcement learning.  The survey includes several
remarks that describe extensions as well as open problems in this
subject.  
\end{abstract}
\maketitle

%
%
%



\section{Introduction}
\label{sec:intro}

In many applications, dynamic models exist only in the
form of a simulator.  Our aim is to provide a survey of a class of algorithms, that use {\em
  only} a model simulator, to solve the two canonical problems of
Control Theory:
\begin{itemize}
\item Design of optimal filter (in the sense of estimation); 
\item Design of optimal control law.
\end{itemize}
In this survey, such simulation-based algorithms are broadly referred to
as {\em controlled interacting particle systems (CIPS)}.  Our research
group's most well
known contribution to CIPS is the feedback particle filter (FPF),
which is also the main focus of this survey. The
FPF algorithm is useful to approximate the optimal (nonlinear) filter.  By
making use of the duality between optimal control and filtering, the
FPF algorithm is extended to approximate the solution of an optimal
control problem.

We begin by describing the high-level idea for the two
problems of optimal filtering and optimal control.

\subsection{CIPS in optimal filtering}

\newP{Mathematical problem} In continuous-time and continuous-space
settings of the problem, the standard model of nonlinear (or stochastic) filtering
is the following It\^o stochastic differential equations
(SDEs):
\begin{subequations}\label{eq:intro-model}
	\begin{align}
		\text{State:}\quad
		\ud X_t &= a(X_t)\ud t + \sigma_B(X_t)\ud B_t,\quad X_0 \sim p_0,
		\label{eq:intro-dyn}
		\\
		\text{Observation:}\quad
		\ud Z_t &= h(X_t)\ud t + \ud W_t,\label{eq:intro-obs}
	\end{align}
\end{subequations}
where $X_t\in\Re^d$ and $Z_t\in\Re^\Zdim$ are the state and
observation, respectively, at time $t$, $p_0$ is the 
probability density function (PDF) at the initial time $t=0$ ($p_0$ is
referred to as the prior density), and $\{B_t\}_{t\geq 0}$,
$\{W_t\}_{t\geq 0}$ are
mutually independent standard Wiener processes (W.P.) taking values in
$\Re^\Bdim$ and $\Re^\Zdim$, respectively. The mappings $a(\cdot)
$, $h(\cdot)$, 
$\sigma_B(\cdot)$, and the density $p_0(\cdot)$ are smooth (continuously
differentiable) functions.  The linear Gaussian model is obtained when
the drift terms $a(\cdot)$, and $h(\cdot)$  are linear functions,
$\sigma_B(\cdot)$ is a constant matrix, and $p_0$ is a Gaussian
density. 

The filtering problem is to compute the conditional PDF of the 
state $X_t$ given the time-history (filtration) of observations up to
time $t$.  The conditional PDF is denoted by $p_t$ and is referred to
as the posterior density.

\medskip

\newP{CIPS algorithm} involves construction of $N$ stochastic processes
$\{X_t^i\in\Re^d: t\geq 0, 1\leq i\leq N\}$ where the $i$-th process (particle)
evolves according to the SDE:
\begin{equation}\label{eq:FPF_intro}
\ud X^i_t =  \underbrace{a(X_t^i) \ud t +  \sigma_B(X_t^i) \ud
  B_t^i}_{i-\text{th copy of
    model}~\eqref{eq:intro-dyn}} \; +\;
\ud U_t^i,\;\; X_0^i\stackrel{\text{i.i.d.}}{\sim} p_0,
\end{equation}
where $U:=\{U_t^i:t\geq 0, 1\leq i\leq N\}$ is referred to as the {\em coupling}
(with $U=0$, the $N$ processes are un-coupled).  The
goal is to  design the coupling $U$ so that the  empirical distribution of the $N$
particles at any time $t$ approximates the posterior $p_t$:
\begin{equation}\label{eq:FPF_MC}
\frac{1}{N} \sum_{i=1}^N f(X_t^i) \approx \int_{\Re^d} f(x) p_t(x)\ud
x, \quad \forall \; f\in C_b(\Re^d),
\end{equation}
where ``$\approx$'' means that the approximation error goes to zero
(in a suitable sense) as $N\to\infty$ ($C_b(\Re^d)$ is the space of continuous and
bounded functions on $\Re^d$).

A key breakthrough, that appeared around 2010, is that $U$ can be
realized as a mean-field type feedback control
law (``mean-field type'' means that the control law
depends also on the statistics of the stochastic process).  Feedback
particle filter (FPF) is one such example of a mean-field type
control law.  In this paper, we describe the 
FPF, relate it to its historical precursor, the ensemble Kalman filter (EnKF)  algorithm, and
summarize the important developments in this area.

For the filtering model~\eqref{eq:intro-model},  the idea of
controlling the particles to approximate the posterior 
appears in the work of
three groups working independently: the first example of such a control
law appears in~\citep{crisan10} using a certain smoothed form of
observations.  The FPF formula appears in~\citep{yang2011mean,yang2011feedback}
and its special case for the linear Gaussian model 
is described in~\citep{reich11,bergemann2012ensemble}.  A comparison of these three
early works can be found in~\citep{pathiraja2020mckean}.
For the discrete-time filtering models, closely related ideas and
algorithms were proposed, also around the same time-frame,
by~\citep{DaumHuang08,MarzoukBayesian,reich2013nonparametric,yang2014continuous}
(see~\citep{spantini2022coupling} for a recent review of this 
literature).

Our early work on FPF was closely inspired by the pioneering
developments in mean-field
games~\citep{huang07large-population,huang06large}.  The topic of
mean-field games and mean-field type optimal control is concerned with
control and
decision problems arising in interacting particle systems. 
Over the past decade, this topic has
grown in significance with theory and applications
described in several
monographs~\citep{bensoussan2013mean,carmona2018probabilistic,gomes2016regularity}.
In the Physics literature, the study of interacting particle systems is a
classical subject~\citep{liggett1985interacting}.  A canonical example
of an interacting particle system is the coupled oscillators model of
Kuramoto~\citep{kuramoto1975self,strogatz2000kuramoto,dorfler2014synchronization}.  Extensions of
the classical Kuramoto model to mean-field games appears  
in~\citep{yin2011synchronization,carmona2020jet} and to FPF is given in~\citep{Tilton_coupled_oscillator_FPF}.

Design of CIPS to approximate the optimal control law is a more recent
development.  The idea is described next.

\subsection{CIPS in optimal control}

\newP{Mathematical problem}
Consider a finite-horizon deterministic
optimal control problem:
\begin{subequations}\label{eq:nonlinear_opt_control_problem}
\begin{align}
\min_{u} \quad J(u) &= 
                                    \int_0^T 
                                    \left(\half |c(x_t)|^2 + \half u_t^\transpose \RU u_t \right)    \ud t  + g(x_T),
         \\
\text{subject to:} \quad \dot{x}_t &= a(x_t) + b(x_t) u_t , \; x_0=x.\label{eq:nonlinear:model-dynamics}
\end{align} 
\end{subequations}
where  $x_t\in\Re^d$ is the state at time $t$ and $u:=\{u_t\in\Re^m:
0\leq t\leq T\}$ is the control input.   The mappings $a(\cdot)$,
$b(\cdot)$, $c(\cdot)$, $g(\cdot)$ are smooth 
functions and $\RU$ is a strictly positive-definite matrix (henceforth
denoted as $\RU\succ 0$).
The linear quadratic (LQ) model is obtained when $a(x) = Ax$, $b(x) =
B$, $c(x) = C x$, and $g(x) = x^\transpose P_T x$.  The infinite-time horizon
($T=\infty$) case is referred to as the linear quadratic regulator
(LQR) problem.

\medskip

\newP{CIPS algorithm} involves construction of $N$ stochastic
processes $\{Y_t^i\in\Re^d:   0\leq t\leq T, 1\leq i\leq N\}$ where
the $i$-th particle evolves according to an SDE
\begin{subequations}\label{eq:CIPS_intro_all}
\begin{equation}\label{eq:CIPS_intro}
\ud Y^i_t =  \underbrace{a(Y_t^i) \ud t + b  (Y_t^i) \ud
  v_t^i}_{i-\text{th copy of
    model}~\eqref{eq:nonlinear:model-dynamics}} \; +\;U_t^i \ud t ,\;\; 0\leq t\leq T,
\end{equation}
where the input $v:=\{v_t^i\in\Re^m:0\leq t\leq T\}$ and the coupling $U:=\{U_t^i\in\Re^d:0\leq t\leq T\}$ are obtained as part of the design.
The goal is to  design $v$ and $U$ so that the  empirical
distribution of the $N$ particles at time $t$ approximates a smooth
density $p_t$ encoding the optimal control law $u_t=\fee^*_t(x_t)$ where
\begin{equation}\label{e:policy}
\fee_t^*(x) =   \RU^{-1}b^\transpose (x) \nabla \log {p}_t(x),\;\; 0\leq t\leq T,
\end{equation}
and $\nabla$ denotes the gradient operator.
In the
infinite-horizon case, a stationary policy is obtained by letting
$T\to\infty$.  
\end{subequations}

The righthand-side of the formula~\eqref{e:policy} is a consequence of the log transformation.
The transformation relates the value function of an optimal control
problem to the posterior density of the dual optimal filtering 
problem~\citep{fleming-1982,mitter2003}.  This manner of converting an optimal
control problem into an optimal filtering problem (and vice-versa) is
referred to as the minimum energy
duality~\citep{hijab1980minimum,mortensen-1968}.  The use of this
duality to express and solve an estimation problem as an optimal control problem
is a standard approach in model predictive
control~\citep[Ch.~4]{rawlings2017model}.  The
CIPS~\eqref{eq:CIPS_intro} comes about from the use of duality in the
opposite direction whereby an optimal control
problem~\eqref{eq:nonlinear_opt_control_problem} is solved using 
a filtering-type algorithm.  Related constructions, based on somewhat
different algorithmic approaches, is an important theme in the
Robotics
literature~\citep{todorov-2007,kappen-2005,kappen-2005-jsm,rawlik2013stochastic,toussaint-2009,hoffman-2017}
(see~\citep{levine-2018} for a recent review).

Both~\eqref{eq:FPF_intro} and~\eqref{eq:CIPS_intro_all} are examples
of a 
``simulation-based'' algorithm because multiple copies---of the
model~\eqref{eq:intro-dyn} and~\eqref{eq:nonlinear:model-dynamics},
respectively---are run in a Monte-Carlo manner.  
{\em The main message of our paper is that \revise{through} a suitable design of
  interactions between simulations---referred to as coupling---yields powerful algorithms for solving optimal filtering and optimal
  control problems}.

\subsection{Relationship to other simulation-based algorithms}

For the two problems of filtering and control, related simulation-based
solution approaches are considered in the data assimilation (DA) and
reinforcement learning (RL) communities, respectively. These relationships
are discussed next.    

\paragraph{1. Data assimilation (DA)} The term ``Data Assimilation''
means assimilating real-time observations (``data'') into models---which
typically exist only as a software code.  The term is used by a
community of researchers working in geophysical and atmospheric
sciences~\citep{van1996data,evensen2006,houtekamer01,reich2015probabilistic}. The
most celebrated application is
weather prediction and forecast.  For the abstract
mathematical model, the nonlinear filter gives the
optimal solution.  In practice, the filter must be approximated in a
computationally tractable form. For this purpose, the EnKF algorithm was first introduced
in~\citep{evensen1994sequential} as an alternative to the extended
Kalman filter (EKF).  In geophysical applications, there are two
issues that adversely affect the
implementation of an extended Kalman filter:
\begin{enumerate}
\item In high-dimensions, it is a challenge to compute the Kalman gain.
  This is because the formula for the Kalman gain is based on the
  solution of a certain differential Riccati equation (DRE).  The
  matrix-valued nature of the DRE means that any algorithm is
$\mathcal{O}(d^2)$ in the dimension $d$ of the state-space. 
\item The model parameters are not explicitly available to write
  down the DRE let alone solve it. This is a concern whenever 
the model exists only in the form of a black-box numerical simulator. 
\end{enumerate}

In an EnKF implementation, $N$ processes are simulated (same
as~\eqref{eq:FPF_intro}). In order to compute the Kalman
gain, the solution of
the DRE at time $t$ is approximated by the empirical covariance of
the ensemble $\{X_t^i\}_{i=1}^N$.  Because an explicit solution of the
DRE is avoided, an EnKF can be implemented 
using only a model simulator.  This property has historically proved
to be an important factor in applications.  
Notably, the EnKF algorithm is a workhorse for
the weather prediction application~\citep{evensen2003ensemble,houtekamer2016review}.  
The computational complexity of the EnKF is
$\mathcal{O}(N d)$ and in high-dimensions, $N$ is chosen to
be much smaller than $d$.    

The historical significance of the FPF is that it represents a
simulation-based solution of the nonlinear filtering
problem~\eqref{eq:intro-model}, for arbitrary types of non-Gaussian
posterior density $p_t$ (under some mild technical conditions).  Moreover, the
EnKF was shown to arise as a special case in the linear Gaussian
setting of the problem.  Like the Kalman filter, the FPF formula has a
``gain times error'' feedback structure which is useful in several ways, e.g.,
to handle additional uncertainty in signal and measurement models. 
For these reasons, FPF can be viewed as a modern extension
to the Kalman filter, a viewpoint stressed in a prior review
paper~\citep{TaghvaeiASME2017}.

For the nonlinear filtering
problem~\eqref{eq:intro-model}, the FPF represents
an alternative solution approach to the sequential
importance sampling-resampling (SIR) particle
filters and its many
variants~\citep{gordon93,bain2009,delmoralbook,doucet09}.  In an SIR
filter, the posterior is approximated as (compare with~\eqref{eq:FPF_MC})
\[
\int_{\Re^d} f(x) p_t(x)\ud
x \approx \sum_{i=1}^N W_t^i f(X_t^i), \quad \forall \; f\in C_b(\Re^d),
\]
where $X_t^i$ is a copy of the hidden state $X_t$ and
$\{W_t^i\}_{i=1}^N$ are the importance  weights obtained from the
Bayes' formula.  In practice, all but a few weights can become very
small---an issue known as particle degeneracy. This issue
is ameliorated using a re-sampling procedure.  The salient feature of
the FPF, compared to the conventional particle filters, is that the
weights are uniform ($=\frac{1}{N}$) by construction.  Because of this difference,  FPF
does not suffer from the particle degeneracy issue and does not
require re-sampling. In several independent
numerical evaluations and comparisons, it has been observed that FPF
exhibits smaller simulation
variance~\citep{berntorp2015,adamtilton_fusion13,yang2013,stano2014}
and better scaling properties with the problem dimension compared to
particle filters~\citep{surace_SIAM_Review,yang2016}.  Some of 
these
analytical and numerical comparisons are 
highlighted in  the paper.

\paragraph{2. Reinforcement learning (RL)} RL is concerned with solving
optimal control problems, such
as~\eqref{eq:nonlinear_opt_control_problem} and its extensions.  All of the standard
choices are treated in the literature: continuous and discrete
state-space and time, deterministic and stochastic dynamics,
discounted and average cost structures, and finite and infinite time-horizon~\citep{bertsi96a,meyn2022control}.  
What makes the RL paradigm so different from optimal control as
formalized by Bellman and Pontryagin in the 1950s 
is that in RL the system identification step is usually avoided.
Instead, the optimal policy is approximated (``learned'') based on
input-output measurements.

In popular media, RL is described as an ``agent'' that learns an
approximately optimal policy based on interactions with the
environment.  Important examples of this idea include advertising,
where there is no scarcity of real-time data.  In the vast majority of
applications we are not so fortunate, which is why successful
implementation usually requires simulation of the physical system for
the purposes of training.   For example, DeepMind's success story with
Go and Chess required weeks of simulation for training on a massive
collection of
super-computers~\citep{mu-zero_2019MasteringAG}.  

These success stories are largely empirical.
In order to better understand the theoretical foundations of RL, there
has been a concerted recent interest, in the Control community, to
revisit the classical linear quadratic (LQ) optimal control
problem~\citep{fazel_global_2018,tu_gap_2019,dean_sample_2020,
  malik_derivative-free_2020,mihailo-2021-tac}.  The two issues 
discussed as part of DA are relevant also to this problem: In
high-dimensions, it is a challenge to solve the Riccati equation, and
typically the model parameters are not explicitly available in RL
settings of the problem.

An outgrowth of this recent work is a class of simulation-based
algorithms where multiple copies of the simulator are run in parallel
to learn and iteratively improve the solution of the DRE.  The CIPS 
algorithm~\eqref{eq:CIPS_intro} has the same structure where the important distinction is
that the simulations are now coupled with a 
coupling term.  We include comparisons
on a benchmark problem to show how coupling helps improve performance
over state-of-the-art.

\subsection{Structure of the paper and outline}

This paper is divided into two parts as follows: 
\begin{itemize}
\item Part I on CIPS for
the optimal filtering problem~\eqref{eq:intro-model}.  It
comprises~\Sec{sec:filtering-background}~-~\Sec{sec:OT-FPF}. 
\item Part II on CIPS for the optimal
control problem~\eqref{eq:nonlinear_opt_control_problem}. It comprises~\Sec{sec:LQR}. 
\end{itemize}

The paper is written so that the key ideas are easily accessible
together with an understanding of the main computational problems and
algorithms for the same.  For example, a reader should to be able to
implement the FPF and EnKF algorithms after reading~\Sec{sec:FPF} and~\Sec{sec:gain-approx}.  
The more theoretical aspects related to optimal transportation theory appear
in a self-contained manner in~\Sec{sec:OT-FPF}.  
The other significant aspect of
this survey is analytical and numerical comparison against competing approaches.
These appear in~\Sec{sec:FPF_comp} for part I where a comparison with the SIR
filter is discussed; and in~\Sec{sec:dual-EnKF-comparison} for part II where a comparison with
RL algorithms for the LQR problem is described.

In writing any survey or review paper, one must make a choice of not only the
topics to include but also the ones to leave out.  Our choice is
guided by our own area of expertise and by the
intended audience in the Control community (where most of our own prior
work
has been published). We have stressed the
interpretation of coupling as a mean-field feedback control law and
highlighted its connection to optimal transportation.  
Both of these are
important research themes in the community with related work on
mean-field optimal control.  The mathematics is most elegant in the
continuous-time settings of the problem which is also the setting of
this paper.  A number of important aspects have not been covered in
detail:   On the theoretical side, the
well-posedness of the mean-field model and justification of the
mean-field limit are both hard mathematical topics.  For a reader
interested in some of these topics, we have included some high level
remarks with references where additional details can be found.   On the
practical side, important issues arise on account of numerical
discretization of the SDEs.  Such numerical aspects have been entirely
left out of this paper.

We make note of two final points: (i) While the paper presents
some relatively novel ideas that are closely inspired by and connected
to the work in mean-field modeling and control, and therefore of
interest to the Control community, these algorithms have older roots
(EnKF) in the DA community.  Along with the discussion in the
Introduction, several remarks are included to highlight these roots
and connections. (ii) While the CIPS algorithms solve some
problems (such as particle degeneracy), they also create new ones.  This informs the structure of
the paper with a dedicated \Sec{sec:gain-approx} on the central
numerical problem of FPF. In particular, the discussion of the
bias-variance trade-off in~\Sec{sec:diff_map} is helpful to
understand some of the key limitations in high dimensions.

\begin{center}
PART I
\end{center}

\section{Background on optimal filtering}
\label{sec:filtering-background}

Consider the filtering problem for the
model~\eqref{eq:intro-model}.  The sigma-algebra (on the time-history) of observations up to time $t$ is denoted by $\mathcal Z_t :=
\sigma(Z_s:  0\le s \le t)$.  The posterior density $p_t$ is defined
as follows:
\[\int_{\Re^d} f(x) p_t(x) \ud x \eqdef \mathbb E[f(X_t)|\clZ_t],\quad \forall \;f\in C_b(\Re^d),
\] 
where the conditional expectation on the righthand-side is 
referred to as the nonlinear filter.  The integral on the
lefthand-side is denoted by $\langle p_t,f\rangle$.

The posterior $p_t$ is optimal in the sense that, among all
$\clZ_t$-measurable random variables, $\langle p_t,f\rangle$ represents 
the best mean-squared error (MSE)~estimate of the random variable $f(X_t)$:
\begin{equation}
\label{eq:mse}
	\langle p_t,f\rangle=\argmin_{S\in \mathcal \clZ_t}~\mathbb E[|f(X_t) -S|^2 ],
\end{equation}
where the notation ``$S\in \mathcal \clZ_t$'' means $S$ is allowed to be
$\clZ_t$-measurable, i.e., an arbitrary measurable function of observations up to
time $t$.  

For the
model~\eqref{eq:intro-model}, the evolution of the posterior $p_t$ is given by the
Kushner-Stratonovich stochastic partial differential equation~\cite[Ch. 5]{xiong2008}.  
In the special linear Gaussian setting of the problem, the equation
admits a finite-dimensional representation given by the
Kalman-Bucy filter. 

\subsection{Linear Gaussian model and the Kalman-Bucy filter} \label{sec:intro-Kalman-filter}

The linear Gaussian model is a special case 
of~\eqref{eq:intro-dyn}-\eqref{eq:intro-obs} and 
takes the following form:
\begin{subequations}\label{eq:model-linear}
	\begin{align}
		\ud X_t &= A X_t + \sigma_B \ud B_t,\quad X_0 \sim \calN(m_0,\Sigma_0), \label{eq:intro-dyn-lin} \\
		\ud Z_t &= H X_t \ud t + \ud W_t,\label{eq:intro-obs-lin}
	\end{align}
\end{subequations}
where $A,H,\sigma_B$ are matrices of appropriate dimensions and the
prior is a Gaussian density with mean $m_0$
and variance $\Sigma_0$.  It is denoted by $\calN(m_0,\Sigma_0)$.

For the linear Gaussian model~\eqref{eq:model-linear}, it can be shown
that the posterior $p_t$ is a Gaussian density.  It is denoted by $\calN(m_t,\Sigma_t)$,
where $m_t$ and $\Sigma_t$ are conditional mean and covariance,
respectively. Their evolution is described by the 
Kalman-Bucy filter~\citep{kalman-bucy}:
\begin{subequations}\label{eq:Kalman-filter}
	\begin{align}
		\ud m_t &= Am_t + \K_t (\ud Z_t - Hm_t \ud t),\quad m_0 \;\text{(given)}\label{eq:intro-Kalman-mean}\\
		\frac{\ud}{\ud t }\Sigma_t &= \Ricc(\Sigma_t),\quad \Sigma_0 \;\text{(given)} \label{eq:intro-Kalman-cov}
	\end{align}
\end{subequations}
where  $\K_t \eqdef \Sigma_t H^\transpose $ is referred to as the Kalman gain, and the Riccati function  
\begin{equation*}
	\Ricc(\Sigma) \eqdef A \Sigma + \Sigma A^\transpose +\Sigma_B - \Sigma H^\transpose H\Sigma
\end{equation*}
with $\Sigma_B \eqdef \sigma_B \sigma_B^\transpose$. 

Apart from the linear Gaussian model, there are very few examples
where the equation for the posterior $p_t$ admits a finite-dimensional
representation~\citep{benevs1981exact}.  In the general setting of the nonlinear
model~\eqref{eq:intro-model} with a non-Gaussian posterior, $p_t$ is
numerically approximated.

\section{Feedback particle filter}
\label{sec:FPF}

Feedback particle filter (FPF) is a numerical algorithm to approximate
the posterior $p_t$ for the filtering model~\eqref{eq:intro-model}.  
Before describing the FPF, it is helpful to consider a simpler static
problem.

\subsection{Intuitive explanation with a simpler example}\label{sec:FPF-simpler-example}
Suppose the state $X$ and the observation $Y$ are
vector-valued random variables of dimension $d$ and $\Zdim$, respectively.  The probability distribution (prior) of $X$ is denoted by ${\sf
  P}_X$ and the joint distribution of $(X,Y)$ is denoted by ${\sf
  P}_{XY}$. For any given function $f\in C_b(\Re^d)$, the problem is
to obtain an MSE.~estimate of the unknown $f(X)$ from a single
observation of $Y$.   
Adapting~\eqref{eq:mse} to the simple case, 
\begin{equation}\label{eq:mse_simple}
	S^*_f(Y) 
	= \argmin_{S_f(\cdot)} \mathbb E[|f(X)-S_f(Y)|^2],
\end{equation}
where on the righthand-side $S_f:\Re^m\to \Re$ is allowed to be an
arbitrary function of the $\Re^m$-valued observation (the sub-script means that the function may
depend also upon $f$).  
The optimal estimator gives the conditional
expectation, i.e., $\mathbb E[f(X)|Y] = S^*_f(Y)$.

\begin{example}[Linear estimation and the update formula for Kalman filter]\label{ex:ex1}
	Consider the case where $f$ is linear, $f(x)=a^\transpose x$, and $S_f(\cdot)$ is
        restricted to be an affine function of its argument: 
	\begin{equation*}
		S_f(y) = u^\transpose y+b,
	\end{equation*}
	where $u \in \mathbb R^m$ and $b \in \mathbb R$ parametrize
        the estimator. 
With such a choice, the optimization problem~\eqref{eq:mse_simple} is
finite-dimensional whose solution is readily obtained as
	\begin{align*}
		S^*_f(Y) =  a^\transpose (\mathbb E[ X] + K(Y-\mathbb E[Y])),
	\end{align*}
	where $K=\Sigma_{XY}\Sigma_Y^{-1}$, $\Sigma_{XY} =
        \mathbb E[(X-\mathbb E[X])(Y-\mathbb E[Y])^\transpose]$, $\Sigma_{Y} =
        \mathbb E[(Y-\mathbb E[Y])(Y-\mathbb E[Y])^\transpose]$, and
        it is assumed that $\Sigma_Y$ is invertible with inverse $\Sigma_Y^{-1}$.  Because
        the vector $a$ is arbitrary, this also shows that the optimal
        linear estimate of $X$ is $\mathbb E[ X] + K(Y-\mathbb
        E[Y])$.  Under the stronger assumption that $X$ and $Y$ are jointly Gaussian, it can be shown that
        this is in fact the optimal estimate of $X$ among all functions
        $S_f(\cdot)$ (not necessarily affine)~\cite[Prop. 3.9]{hajek2015random}.  Therefore, in
        the Gaussian case
	\begin{equation*}
		\mathbb E[X|Y] = \mathbb E[ X] + K(Y-\mathbb E[Y]). 
	\end{equation*}
The righthand-side is the update formula for the discrete-time
Kalman filter. Note that the interpretation of the formula as the
conditional expectation works
only in the Gaussian case.  In general, the formula gives only
the best linear estimator.   

\end{example}
The example above illustrates the special and important case of
obtaining optimal linear estimators. The question is how to extend the
procedure to the nonlinear setting, i.e., the setting where both the
function $f(\cdot)$ and the estimator $S_f(\cdot)$ are allowed to be
nonlinear functions of their arguments.  This is achieved through the
concept of CIPS whose construction proceeds in two steps: 

\medskip

\newP{Step 1} Let $\bar X_0$ be an independent copy of $X$.  Design a
control $U$ such that, upon setting $\bar X_1=\bar X_0 + U$, 
\begin{equation*}
	S_f^*(Y) = \mathbb E[f(\bar X_1)|Y],\quad \forall\; f \in C_b(\Re^d),
\end{equation*} 
Note that
the control is not allowed to depend on the function $f$. It is
designed to give the best estimate for any choice of function $f$. 
It
is not yet clear that such a control exists. But for now, let us
assume that it exists and moreover takes the form $U = u(\bar X_0,Y)$.
(Typically, the mapping $u(\cdot,\cdot)$ is designed to be a
deterministic function but may in general also be random.)

\medskip

\newP{Step 2} Generate $N$ independent samples (particles) $\{
X^1_0,\ldots, X^N_0\}$ from ${\sf P}_X$, update each particle according to   
\begin{equation*}
	X^i_1 = X^i_0 + u(X^i_0,Y),\quad i=1,2,\hdots,N,
\end{equation*}
and form a Monte-Carlo approximation of the estimate: 
\begin{equation*}
	S_f^*(Y) \approx \frac{1}{N}\sum_{i=1}^N f(X^i_1).
\end{equation*}

\begin{example}[CIPS and the update formula for EnKF] 
\label{ex:ex2}
Continuing with Ex.~\ref{ex:ex1} where ${\sf
    P}_{XY}$ is assumed to be Gaussian, two
  formulae are described for the transformation $\bar X_0\mapsto \bar X_1$.  The first of these
  formulae is based on optimal transportation
  theory.  The second formula is based on the 
  perturbed form of the discrete-time EnKF algorithm.  
\begin{itemize}
\item Optimal transport formula is given by a deterministic affine
  mapping 
\[
\bar X_1 = A(\bar X_0 - \mathbb E[\bar X_0]) + K(Y - \mathbb E[Y]) +  \mathbb E[\bar X_0],
\]
where $A$ is the unique such symmetric positive-definite solution to a
Lyapunov equation
\[
		A \Sigma_X A = \Sigma_X - \Sigma_{XY} \Sigma_Y^{-1} \Sigma_{YX}. 
\]
\item Perturbed EnKF formula.  Let $(\bar X_0, \bar Y_0)$ be an
  independent copy of $(X,Y)$ then 
\[
\bar X_1 = \bar X_0 + K(Y - \bar Y_0),  
\]
where the formula for $K$ is same as in Ex.~\ref{ex:ex1}.  
\end{itemize}   
It is readily verified that, in either case, $\bar{X}_1$ is a Gaussian
random variable whose conditional mean and variance equals the
conditional mean and variance of $X$.   

We defer the details on how these formulae came about
to~\Sec{sec:OT-FPF-simpler-example} instead remarking here 
on several features which apply also to more general settings:
\begin{enumerate}
\item The transformation $\bar X_0\mapsto \bar X_1$ is not unique. 
\item Both the transformations are of ``mean-field type'' whereby the
  transformation depends also on statistics, e.g., $\mathbb E[X]$ and
  $\mathbb E[Y]$, of $(X,Y)$.
\item In the optimal transport formula, $u(\cdot,\cdot)$ is
  a deterministic function.  In the EnKF formula, $u(x,y) = K (y -
  \bar{Y}_0)$ is a random map because $Y_0$ is a random variable.  
\end{enumerate}

\end{example}

These considerations provide the background for the feedback particle filter algorithm
which is described next.

\subsection{Feedback particle filter}

Just like the static example, the construction of FPF proceeds in two steps.

\medskip

\newP{Step 1} Construct a stochastic process, denoted by
$\bar{X}=\{\bar{X}_t\}_{t\geq 0}$, according to a controlled SDE:
\begin{equation}\label{eq:Xbar-u-K}
	\ud \bar X_t = a(\bar X_t)\ud t + \sigma_B(\bar X_t)\ud B_t + u_t \ud t + \mathsf K_t \ud Z_t,\quad \bar X_0 \sim p_0,
\end{equation}
	where the controls $u_t$ and $\mathsf  K_t$ are designed so that
        the conditional density of $\bar X_t$ equals the
        posterior density $p_t$.

\medskip
	
\newP{Step 2} Simulate $N$ stochastic processes, denoted by
$X^i=\{X^i_t\}_{t\geq 0}$ for $i=1,2,\hdots,N$, according to~\eqref{eq:Xbar-u-K}.  

\medskip

The two steps are summarized below:
\begin{equation*}
	\underbrace{\langle p_t,f\rangle\overset{\text{Step
				1}}{=}\mathbb{E}[f(\bar{X}_t)|\clZ_t]}_{\text{exactness condition}}
	\overset{\text{Step 2}}{\approx} \frac{1}{N}\sum_{i=1}^N f(X^i_t).\label{eq:exactness}
\end{equation*}
The exactness condition refers to the fact that $\bar{X}_t$ has the
same conditional density as $X_t$.  The $N$ processes
$\{X^i\}_{i=1}^N$ are referred to as particles.  

\medskip

At this point, the first of these two steps appears to be
aspirational.  Even in the case of the static example, it is not at
all clear that the function $u(\cdot,\cdot)$ exists in the general
non-Gaussian case,
and even if it does, it can be computed in a tractable manner.  The
case of the stochastic process where $u_t$ and $\K_t$ are allowed to
be measurable with respect to the past values of observations $Z$ and
state $\bar{X}$ appears, at the first glance, to be entirely hopeless.

The surprising (at least at the time of its discovery) breakthrough of
the FPF is that the control terms $u_t$ and $K_t$ are given by a 
simple feedback control law where the computation reduces to solving a
linear Poisson equation at each time-step.  

\medskip

\newP{FPF}
The process $\bar{X}$ is defined according to the SDE
\begin{align}
	 \ud \bar{X}_t & = \underbrace{a(\bar{X}_t) \ud t  +
  \sigma_B(\bar{X}_t) \ud \bar{B}_t}_{\text{copy of
  model}~\eqref{eq:intro-dyn}} \nonumber \\
& \qquad \qquad + \underbrace{\K_t(\bar{X}_t) \circ (\ud Z_t -
		\frac{h(\bar{X}_t) + \bar{h}_t}{2}\ud
  t)}_{\text{FPF feedback control law}},\quad \bar{X}_0\sim p_0
	\label{eq:intro-FPF-Xbar}
\end{align}  
where 
$\{\bar{B}_t\}_{t\geq 0}$ is a copy of the process noise $\{B_t\}_{t\geq 0}$, and
$\bar{h}_t := \mathbb{E}[h(\bar{X}_t)|\mathcal{Z}_t]$. 
The $\circ$ indicates that the SDE is expressed in its Stratonovich
form.  At any fixed time $t$, the gain $\K_t(\cdot)$ is a $d\times
\Zdim$ matrix-valued function obtained by solving $m$ partial differential equations: for $j=1,2,\ldots \Zdim$, the $j$-th column $\K^{(j)}_t := \nabla \phi^{(j)}$ where
$\phi^{(j)}$ is the solution of the Poisson equation:
\begin{equation}
	\label{eq:intro-Poisson}
 -\frac{1}{\rho (x)}\nabla \cdot (\rho (x)
	\nabla \phi^{(j)}(x) ) = (h^{(j)}(x)-\bar{h}^{(j)}),\quad x\in\Re^d
\end{equation}
where the density $\rho=\bar p_t$
(the conditional density of
$\bar{X}_t$ at time $t$),  $h^{(j)}$ is the $j$-th component of the observation function $h$, $\bar{h}^{(j)}=\int {h}^{(j)}(x)\rho(x)\ud x$, 
and $\nabla$ and $\nabla \cdot $ denote the gradient and the
divergence operators, respectively. For a succinct presentation, the functions $\{\phi^{(j)}\}_{j=1}^\Zdim$ are collected to form the vector-valued function $\phi = [\phi^{(1)},\ldots,\phi^{(\Zdim)}] $. With such a notation, the gain function $\K_t$ is the Jacobian $ \nabla \phi = [\nabla \phi^{(1)},\ldots,\nabla \phi^{(\Zdim)}]$.

The process $\bar X$ is an example of a mean-field process because its
evolution depends upon its own statistics.  An SDE of this
type is called a McKean-Vlasov SDE or a mean-field
SDE. Accordingly,~\eqref{eq:intro-FPF-Xbar} is referred to as the mean-field FPF.

The main result, first proved in~\cite{yang2013}, is that the mean-field
process thus defined 
is exact. 
\begin{theorem}[Thm~3.3, \cite{yang2013}]
Consider the filtering model~\eqref{eq:intro-model}. Suppose
$\{p_t\}_{t\geq 0}$ denotes the conditional density of the process
$\{X_t\}_{t\geq 0}$.  Suppose the mean-field process
$\{\bar{X}_t\}_{t\geq 0}$ defined
by~\eqref{eq:intro-FPF-Xbar}-\eqref{eq:intro-Poisson} is well-posed
with conditional density denoted by $\{\bar{p}_t\}_{t\geq 0}$.  
Then, provided $\bar p_0 = p_0$,
\begin{equation*}
	\bar p_t = p_t,\quad \forall \; t>0.
\end{equation*}
\label{thm:exactness}
\end{theorem}

\begin{remark}[Well-posedness and Poincar\'e inequality] The well-posedness
of~\eqref{eq:intro-FPF-Xbar}-\eqref{eq:intro-Poisson} means that a
strong solution $\bar{X}$ exists with a well-defined density $\{\bar
p_t\}_{t\geq 0}$.  To show well-posedness, apart from the standard
Lipschitz condition on the drift terms $a(\cdot)$ and
$\sigma_B(\cdot)$, the main technical condition is that the posterior density $p_t$ (of $X_t$) satisfies
the Poincar\'e inequality (PI), and $\int |h(x)|^2 p_t(x) \ud x < \infty$~\cite[Theorem 2.2]{laugesen15}.  (A probability density
$\rho=e^{-V}$ satisfies the PI if $x^\transpose \nabla V(x)\geq
\alpha |x|$ for $|x|\geq R$ where $\alpha$ and $R$ are positive constants~\citep[Cor. 1.6]{bakry08}. This condition is true, e.g., whenever
$\rho$ has a Gaussian tail.)  An explanation of the relevance of the PI
for the well-posedness (existence,
uniqueness, and regularity) of the solution $\phi$ of the Poisson
equation~\eqref{eq:intro-Poisson} is deferred to~\Sec{sec:gain-approx},
where algorithms for its approximation are also described. Once a solution
$\phi$ of the Poisson equation is obtained together with necessary
apriori estimates, well posedness of $\bar{X}$ follows
from the standard theory of mean-field
SDEs~\citep{carmona2018probabilistic}.  Although the general case
remains open, it has been possible to prove the PI under
certain additional conditions on the filtering model~\eqref{eq:intro-model}
	~\cite[Lemma 5.1]{pathiraja2020mckean},~\cite[Prop
        2.1]{laugesen15}. 
\end{remark}

We next describe the finite-$N$ algorithm which is how the FPF is implemented in
practice.  

\medskip

\newP{CIPS}
The particles $\{X^i_t:t\geq 0,1\leq i\leq N \}$ evolve according to:
\begin{equation}
\begin{aligned}
	\ud X^i_t &= a(X^i_t) \ud t + \sigma(X_t^i) \ud B^i_t \\
&\qquad +{\K^{(N)}_t(X^i_t) \circ (\ud Z_t -
		\frac{h(X^i_t) + {h}^{(N)}_t}{2}\ud t)},\\
  X^i_0&\overset{\text{i.i.d}}{\sim} p_0,\quad\quad i=1,\ldots N,
\end{aligned}  
		\label{eq:intro-FPF-finite-N}
\end{equation}
where $\{B^i_t\}_{t\geq}$, for $i=1,2,\hdots,N$, are mutually independent W.P.,
${h}^{(N)}_t:=N^{-1} \sum_{i=1}^N h(X^i_t)$, and $\K^{(N)}_t$
is the output of an algorithm that is used to approximates the solution to the Poisson equation~\eqref{eq:intro-Poisson}: 
\begin{equation*}
\K^{(N)}_t := \text{Algorithm}(\{X^i_t\}_{i=1}^N;h).
	\label{eq:intro-gain-func-approx} 
\end{equation*}
The notation is suggestive of the fact that algorithm is adapted
to the ensemble $\{X^i_t\}_{i=1}^N$ and the function $h$; the density
$\bar p_t$ is not known in an explicit form.  Before describing the
algorithms for gain function approximation in (the following)
\Sec{sec:gain-approx}, we discuss the linear Gaussian case.

The main computational challenge to simulate the finite-$N$ FPF~\eqref{eq:intro-FPF-finite-N}
 is the computation of the gain function.   The difficulty arises
 because, for a general nonlinear observation
function $h$ and a non-Gaussian density $\rho$, 
there are no known closed-form solutions of the Poisson
equation~\eqref{eq:intro-Poisson}.  
In the 
linear Gaussian special case, with a linear observation
function $h(x)=Hx$ and a Gaussian
density, the Poisson equation admits an explicit solution whereby the
gain function is given by the Kalman gain: 

\begin{proposition}[Lem.~3.4, \cite{yang2013}]\label{prop:K-Gaussian}
	Consider the Poisson equation~\eqref{eq:intro-Poisson}.
        Suppose   
        $\rho$ is a  Gaussian density ${\cal N}(m,\Sigma)$ and
        $h(x)=Hx$. Then its unique solution is given by:
	\begin{equation*}
		\phi(x) = (H \Sigma) (x-m),\quad x\in\Re^d.
	\end{equation*}
Consequently, the gain function $\nabla \phi(x)=\Sigma H^\transpose$ is the
Kalman gain. 
\end{proposition}

Using the Kalman
gain, the FPF algorithm simplifies to a square-root form of the
ensemble Kalman filter (EnKF) algorithm. This is described next.

\subsection{Ensemble Kalman filter}

In the linear Gaussian case, upon replacing the gain function with the
Kalman gain, the mean-field
        FPF~\eqref{eq:intro-FPF-Xbar} is the It\^o-SDE
\begin{equation}\label{eq:intro-FPF-Xbar-lin}
	\ud \bar{X}_t = A \bar{X}_t \ud t + \sigma_B \ud \bar{B}_t + \bar{\Sigma}_tH^\transpose  (\ud Z_t - \frac{H\bar{X}_t+H\bar{m}_t}{2}\ud t),
\end{equation}  
where 
\begin{align*}
\bar{m}_t&=\mathbb{E}[\bar{X}_t|\clZ_t],\\
\bar{\Sigma}_t&= \mathbb{E}[(\bar{X}_t-\bar{m}_t) (\bar{X}_t-\bar{m}_t)^\transpose|\clZ_t].
\end{align*}
As a corollary of~\Thm{thm:exactness}, the mean-field
process $\bar X$ is exact which, in the linear
Gaussian case, means that the conditional density of $\bar X_t$ is
Gaussian whose mean $\bar{m}_t$ and
the covariance matrix $\bar{\Sigma}_t$ evolve according to the Kalman
filter~\eqref{eq:Kalman-filter}.  A direct proof
showing~\eqref{eq:intro-FPF-Xbar-lin} is exact appears in~\Sec{sec:LG-OT}.

The finite-$N$ FPF is obtained as follows:
\begin{subequations}\label{eq:intro-FPF-finite-N-lin-overall}
\begin{equation}
	\ud X^i_t = AX^i_t \ud t + \sigma_B \ud B^i_t +{\Sigma^{(N)}_t H^\transpose   (\ud Z_t -
		\frac{HX^i_t + H m^{(N)}_t}{2}\ud t)}, \quad\label{eq:intro-FPF-finite-N-lin}
\end{equation}
where the mean-field terms in~\eqref{eq:intro-FPF-Xbar-lin} are approximated empirically as follows:
\begin{align}	\label{eq:empr_app_mean}
	m^{(N)}_t&:=\frac{1}{N}\sum_{j=1}^N X^i_t,\\ \Sigma^{(N)}_t
	&:=\frac{1}{N-1}\sum_{j=1}^N (X^i_t-m^{(N)}_t)(X^i_t-m^{(N)}_t)^\transpose. 
	\label{eq:empr_app_var}
\end{align}
\end{subequations}
The linear Gaussian FPF~\eqref{eq:intro-FPF-finite-N-lin-overall} is identical to the
square-root form of the ensemble
Kalman filter~\cite[Eq. 3.3]{bergemann2012ensemble}.

\begin{remark}[Historical context for EnKF]
The EnKF algorithm was first introduced
in~\cite{evensen1994sequential}, in the discrete-time setting of the
filtering problem.  At the time, the algorithm was introduced as an
alternative to the extended Kalman filter (EKF).  As already mentioned
in~\Sec{sec:intro}, a major reason for using an EnKF is that, unlike
EKF, it does not require an explicit solution of the DRE~\citep{van1996data,burgers1998analysis,houtekamer1998data}.  
Since its
introduction, a number of distinct types of EnKF algorithms have
appeared in the literature.  Amongst these, the most well-known types
are as follows: (i) EnKF based on perturbed observation~\citep{evensen2003ensemble};
  and (ii) The square root EnKF~\citep{anderson2001ensemble,whitaker2002ensemble,bishop2001adaptive}. 
The details for these algorithms can be found
in~\cite[Ch. 6-7]{reich2015probabilistic}.       
The two aforementioned types of the EnKF
algorithm have also been extended to the continuous-time
setting~\citep{bergemann2012ensemble}.  In these settings, the EnKF is
usually referred to as the ensemble Kalman-Bucy filter (EnKBF). 
A review of the EnKBF algorithm and its connection to the FPF
algorithm can be found in~\citep{TaghvaeiASME2017}. The EnKBF
algorithm and the linear FPF admits several extensions: (i) EnKBF with perturbed
observation~\citep{bergemann2012ensemble}~\citep{delmoral2016stability};
(ii) Stochastic linear FPF~\citep[Eq. (26)]{yang2016} which is same as
the square root EnKBF~\citep{bergemann2012ensemble};(iii)
Deterministic linear
FPF~\citep[Eq. (15)]{AmirACC2016}~\citep{jana2016stability}.  EnKF was recently extended to the case with correlated observation noise~\citep{ertel2022analysis}.  An
excellent recent survey on this topic appears
in~\cite{calvello2022ensemble}. 
\end{remark}

\begin{remark}[Current research on EnKF]\label{rem:EnKF-error}
Error analysis of the EnKF algorithm remains an active area of
research.  For the discrete-time EnKF algorithm, these results appear
in~\citep{gland2009,mandel2011convergence,tong2016nonlinear,stuart2014stability,kwiatkowski2015convergence}.
The analysis for continuous-time EnKF is more
recent~\citep{delmoral2016stability,bishop2018stability,taghvaei2018error,delmoral2017stability,jana2016stability,bishop2020mathematical,chen2021general}.
Typically,
one is interested in
obtaining a uniform error bound as follows:
\begin{equation}\label{eq:unif-error-EnKF}
\mathbb E[\|\mN_t - m_t\|^2]+\mathbb E[\|\SigN_t - \Sigma_t\|^2]\leq \frac{C}{\sqrt{N}},
\end{equation}
where $(m_t,\Sigma_t)$ are the solutions of the Kalman filter~\eqref{eq:Kalman-filter}
and $(\mN_t, \SigN_t)$ are obtained from simulating an EnKF; and $C>0$
is a time-independent constant. 
In the most recent
result~\citep{bishop2020mathematical}, \eqref{eq:unif-error-EnKF} is shown under the
assumption that $H^\transpose H$ is a positive-definite matrix. It is
expected that \eqref{eq:unif-error-EnKF} also holds under the weaker condition of the pair
$(A,H)$ being detectable, which is the condition for the stability of
the Kalman filter.  However, a complete resolution is still open.  A
comprehensive review of  recent developments in this area can be found
in~\cite{bishop2020mathematical}. 
\end{remark}

\subsection{Comparison with importance sampling}
\label{sec:FPF_comp}

In this section, we provide an analytical comparison of the FPF with the
importance sampling-based particle filter.  For this purpose, 
consider a parameter estimation example with a fully observed
model as follows:
\begin{equation}
	\begin{aligned}
		\ud X_t &= 0,\quad\quad X_0 \sim \mathcal{N}(0,\sigma_0^2 I_d)=p_0,\\
		\ud Z_t &= X_t \ud t +  \sigma_w \ud W_t,
	\end{aligned}
	\label{eq:filter-example}
\end{equation}
where the time $t\in[0,1]$, $\sigma_W,\sigma_0>0$, and $I_d$ is the $d\times
d$ identity matrix. 
The posterior $p_1$ at time $t=1$ is a
Gaussian $\NN(m_1,\Sigma_1)$ with $m_1= \frac{\sigma_0^2}{\sigma_0^2
	+ \sigma_W^2}Z_1$ and $\Sigma_1
=\frac{\sigma_0^2\sigma_w^2}{\sigma_0^2 + \sigma_w^2} I_d$.

Let $\{X^i_0\}_{i=1}^N$ be $N$ i.i.d samples from the prior $p_0$.  The importance
sampling-based particle filter yields an empirical approximation of the
posterior $p_1$ as follows: 
\begin{equation}\label{eq:PF-Est} 
	\pi^{(N)}_\text{PF} (f) := \sum_{i=1}^N W_1^i f(X^i_0),\quad W_1^i = \frac{e^{-\frac{|Z_1-X^i_0|^2}{2\sigma_w^2}}}{\sum_{i=1}^N e^{-\frac{|Z_1-X^i_0|^2}{2\sigma_w^2}}}.
\end{equation}
In contrast, given the initial samples $\{X^i_0\}_{i=1}^N$, the FPF approximates
the posterior by implementing a feedback control law as follows:
\begin{equation}\label{eq:EnKF-est}
	\pi^{(N)}_{\text{FPF}} (f) := \frac{1}{N}\sum_{i=1}^N f(X^i_1),\; \ud X^i_t = \frac{\SigN_t}{\sigma_w^2}(\ud Z_t - \frac{X^i_t + \mN_t}{2} \ud t), 
\end{equation}
where the  mean $\mN_t$ and covariance $\SigN_t$ are
empirically approximated using~\eqref{eq:empr_app_mean}
and~\eqref{eq:empr_app_var}, respectively.

The MSE in estimating the conditional expectation of a given
function $f$ is defined as follows:
\begin{equation*}
	\text{MSE}_*(f) := \mathbb{E}[|\pi_*^{(N)}(f) -  \langle
        p_1,f \rangle|^2],
\end{equation*}
where the subscript $*$ is either the $\text{PF}$ or the
$\text{FPF}$.  

\begin{figure*}[t]
	\centering
	\begin{tabular}{cc}
		\subfigure[]{\includegraphics[width = 0.4\hsize,trim=1cm 0cm 1cm 0cm]{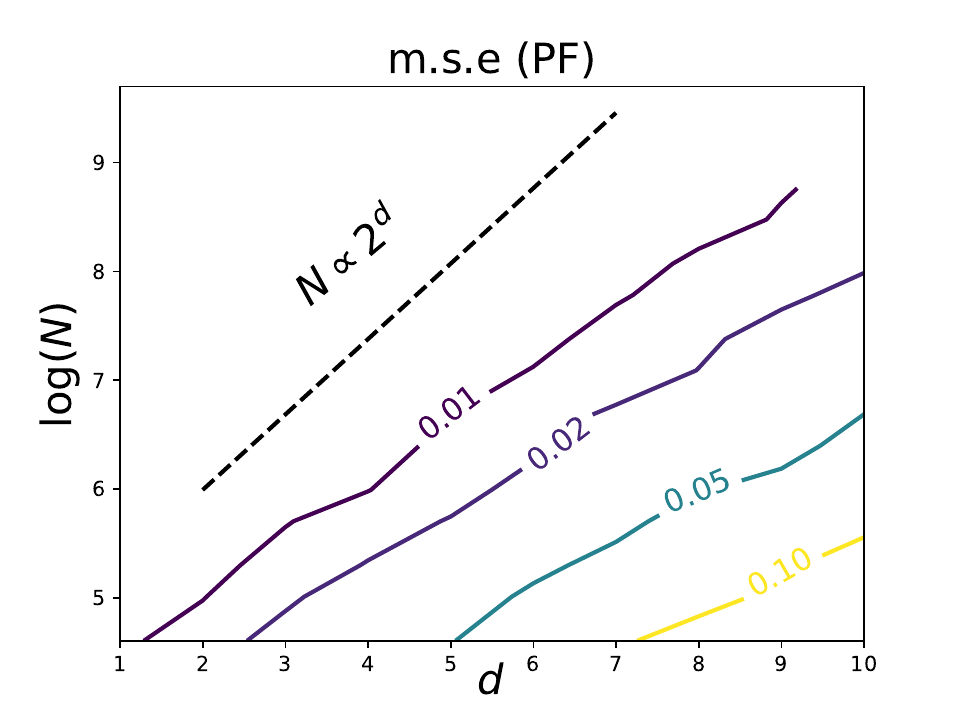}\label{fig:error-PF}} \hspace{10pt}&\hspace{10pt}
		\subfigure[]{\includegraphics[width = 0.4\hsize,trim=1cm 0cm 1cm 0cm]{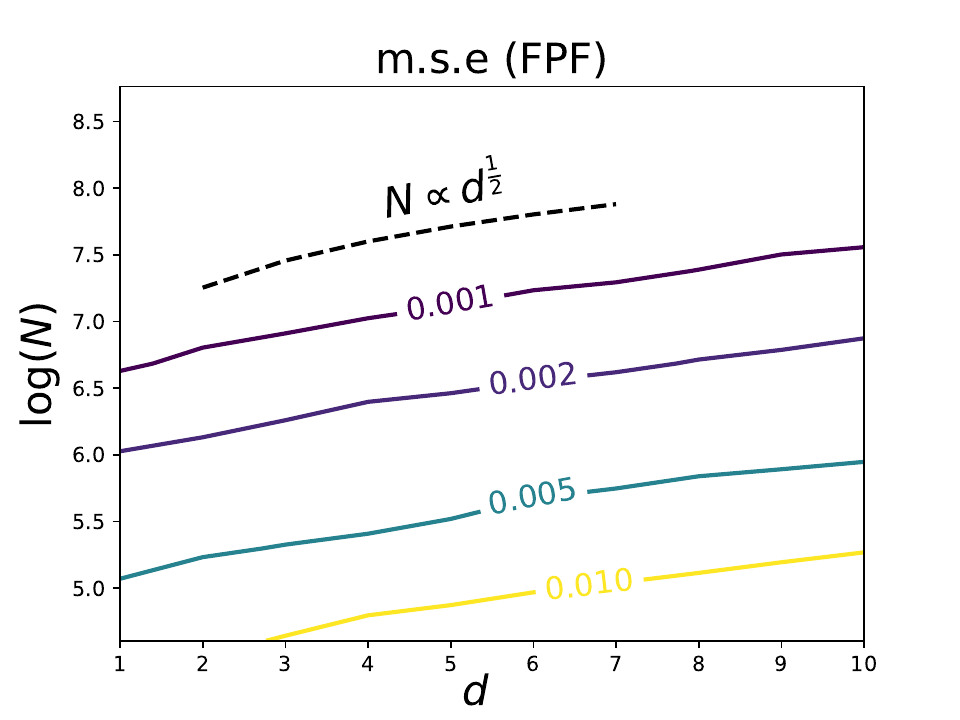}\label{fig:error-FPF}}
	\end{tabular}
	\caption{Numerical comparison for the filtering
          model~\eqref{eq:filter-example}.  Level sets of the MSE.~using: (a) importance
		sampling-based algorithm~\eqref{eq:PF-Est} and (b) the
                FPF~\eqref{eq:EnKF-est}.  As the state dimension $d$
                grows, in order to have same performance (MSE), the
                number of particles $N$ must increase as $2^d$ for~\eqref{eq:PF-Est} while they increase as $d^\half$
                for~\eqref{eq:EnKF-est}.} 
	\label{fig:error-PF-FPF}
\end{figure*}

For $f(x) = \frac{1}{\sqrt{d}}1^\transpose  x$, a numerically computed plot
of the level-sets of MSE, as a function of $N$ and $d$, is depicted in Figure~\ref{fig:error-PF-FPF}-(a)-(b).
The expectation is approximated by averaging over $M=1000$ independent
simulations.  It is observed that, in order to have the same error,
the importance sampling-based approach requires the number of samples
$N$ to grow exponentially with the dimension $d$, whereas the growth
using the FPF for this numerical example is $O(d^\half)$.  This conclusion is consistent
with other numerical studies reported in the literature~\citep{surace_SIAM_Review,stano2014,berntorp2015}.

For the purposes of the analysis, a modified form of the particle
filter is considered whereby the denominator is replaced by its exact
form: 
\begin{equation}
	\pi^{(N)}_{\overline{\text{PF}}} (f) :=  
        \sum_{i=1}^N \Wbar_1^i f(X^i_0),\quad \Wbar_1^i = \frac{e^{-\frac{|Z_1-X^i_0|^2}{2\sigma_w^2}}}{N\mathbb{E}[e^{-\frac{|Z_1-X_0|^2}{2\sigma_w^2}}|\clZ_1]}.
	\label{eq:PF-Est-bar}
\end{equation}

\medskip

\begin{proposition}[Prop.~4 in~\citep{taghvaei2020optimal}]\label{prop:importance-sampling}
	Consider the filtering problem~\eqref{eq:filter-example} with
	state dimension $d$.  
	Suppose
	$\sigma_0=\sigma_w=\sigma>0$ and $f(x)=a^\transpose x$ where
	$a\in\Re^d$ with $|a|=1$.  Then:
	\begin{enumerate}
		\item The MSE.~for the modified 
		importance sampling estimator~\eqref{eq:PF-Est-bar} is given by 
		\begin{equation*}
			\text{MSE}_{\overline{\text{PF}}}(f) = \frac{\sigma^2}{N}\left(3(2^d) - \frac{1}{2}\right) \geq \frac{\sigma^2}{N}2^{d+1}.
		\end{equation*}
		\item The MSE for the FPF estimator~\eqref{eq:EnKF-est} is bounded as
		\begin{equation}\label{eq:FPF-mse-bound}
			\text{MSE}_{{\text{FPF}}}(f) \leq  \frac{\sigma^2}{N} (3d^2+2d).
		\end{equation}
	\end{enumerate}
\end{proposition}

\medskip

\begin{remark}[Curse of
	Dimensionality (CoD)]
	In the limit as $d\to\infty$, the performance of the importance
	sampling-based particle filters is studied in the
	literature~\citep{bickel2008sharp,bengtsson08,snyder2008obstacles,rebeschini2015can}. The
	main focus of these studies is on the particle degeneracy
	(or the weight collapse) issue:  it
	is shown that if $\frac{\log N \log d}{d}\to 0$ then the largest
	weight $\max_{1\le i \le N} W_t^i \to 1$ in probability.  Consequently, in order to prevent the
	weight collapse,
	the number of particles must grow exponentially with the
	dimension. This phenomenon is referred to as the curse of
	dimensionality for the particle filters.  In contrast, the
        weights in an FPF are uniform by design
        (see~\eqref{eq:EnKF-est}).  Therefore, the FPF does not suffer
        from the weight collapse issue and, in particular, does not require
        resampling.  A complete comparison of the two types of
        particle filters remains open
        (see~\citep{abedi2022unification} for recent progress on this
        topic).  
\end{remark}

\begin{remark}[Scaling with the dimension]
	The scaling with dimension depicted
        in~\Fig{fig:error-PF-FPF}~(b) suggests that the $O(d^2)$ bound
        in~\eqref{eq:FPF-mse-bound} is loose.  This is the case because, in deriving
	the bound, the inequality 
	$\|\cdot\|_2\leq \|\cdot\|_F$ is used, where
          $\|\cdot\|$ and $\|\cdot\|_F$ denote the induced and
          Frobenius norms, respectively~\citep[Appendix E]{taghvaei2020optimal}.  The inequality is loose
	particularly so as the dimension grows.  Also, it is observed that the MSE for the particle filter grows slightly slower than the lower-bound $2^d$. This is because the lower-bound is obtained for the modified particle filter~\eqref{eq:PF-Est-bar}, while the MSE is numerically evaluated for the standard particle filter~\eqref{eq:PF-Est}. The correlation between the numerator and denominator in~\eqref{eq:PF-Est} reduces the MSE. 
\end{remark}

\subsection{Extensions of FPF}

In deriving the FPF, the main modeling assumption is the nature of
observation model~\eqref{eq:intro-obs}.  (Such a model is referred to
as the white noise observation model.)  In several follow on works,
the basic FPF is extended to handle more general types of models
for the state process.  These extensions are briefly described next.

\paragraph{1) FPF on Riemannian manifolds}The feedback control form of
the FPF formula~\eqref{eq:intro-FPF-Xbar} holds not only for the
Euclidean state-space but also for the cases where the state
$\{X_t\}_{t\geq 0}$ evolves on a Riemannian manifold,
  such as the matrix Lie groups.  These extensions are described
  in~\citep{zhang2016feedback,zhang2016attitude,zhang2017attitude,zhang2017feedback}.
  In these papers, the FPF is shown to provide an intrinsic
  description of the filter that automatically satisfies the geometric
  constraints of the manifold.  The gain is expressed as
  $\text{grad}\;\phi$ and obtained as a solution of the Poisson
  equation.  It is shown that the gain is also intrinsic that
  furthermore does not depend upon the choice of the Riemannian metric.  For the special case when the manifold
  is a matrix Lie group, explicit formulae for the filter are derived,
  using the matrix coordinates. Filters for two example problems are
  presented: the attitude estimation problem on $SO(3)$ and the robot
  localization problem in $SE(3)$.  Comparisons are also provided
  between the FPF and popular algorithms for attitude estimation,
  namely the multiplicative EKF, the invariant
  EKF, the unscented quaternion estimator, the invariant ensemble
  Kalman filter, and the bootstrap particle filter.  Specifically,
  under a certain assumption of a ``concentrated distribution'', the
  evolution equations for the mean and the covariance are shown to be
  identical to the left invariant EKF algorithm.

\paragraph{2) FPF on discrete state-space}
In~\cite{yang2015feedback}, FPF is extended to the filtering problem
where the hidden state $\{X_t\}_{t\geq 0}$ is a continuous-time Markov process that
evolves on a finite state-space.  (For this model, the optimal
nonlinear filter
is called the Wonham filter.) A standard algorithm to simulate a Markov
process is based on the use of Poisson counters to simulate transitions between discrete states.  In order to
define the process $\bar{X}$, a control process $U$ is introduced that
serves to modulate
the rates of these counters based on causal observations of data $Z$.  An
explicit formula for the FPF feedback control law is derived and shown
to be exact.  Similar
to~\eqref{eq:intro-FPF-Xbar}, the
formula is in the form of ``gain times error'' where the gain is
now obtained by solving a certain linear matrix problem.  The linear
matrix problem is the finite state-space  counterpart of the Poisson
equation~\eqref{eq:intro-Poisson}.

\paragraph{3) FPF with data association and model uncertainty}In applications
such as multiple target tracking, the filtering problem often
involves additional uncertainties in the state
model~\eqref{eq:intro-dyn} and the observation
model~\eqref{eq:intro-obs}.  In the classical linear Gaussian
settings, algorithms based on the Kalman filter have been
developed to provide a solution to these problems. 
These algorithms are referred to as the interacting
multiple model (IMM) filter~\citep{blom2013continuous} and the probabilistic
data association (PDA) filter~\citep{bar2009probabilistic}.  In the PDA
filter, the Kalman gain is allowed to vary based on an estimate of the
instantaneous uncertainty in the observations. In the IMM filter,
multiple Kalman filters are run in parallel and their outputs combined
to form an estimate. 

Like the Kalman filter, the FPF is easily extended to handle
additional uncertainties in the observation and signal models:  These
extensions, namely, the probabilistic data association
(PDA)-FPF and the interacting multiple model (IMM)-FPF are derived in our
prior works~\citep{yang2012joint,yang2013interacting,yang2018probabilistic}.  Structurally,
the FPF based implementations are similar to the classical
algorithms based on the Kalman filter. In the linear
Gaussian settings, the equations for the mean and the variance of
the FPF-based filters evolve
according the classical PDA and IMM filters.

\paragraph{4) Collective inference FPF}The term ``collective
inference'' is used to describe filtering problems with a large number
of aggregate and anonymized
data~\citep{sheldon2011collective,singh2020incremental}.  Some of
these problems have gained in importance recently because of COVID-19.   Indeed, the
spread of COVID-19 involves dynamically evolving hidden processes (e.g.,
number of infected, number of asymptomatic etc..) that must be
deduced from noisy and partially observed data (e.g., number of tested
positive, number of deaths, number of hospitalized etc.).  
In carrying out data assimilation for such problems, one typically only has
aggregate observations.  For
example, while the number of daily tested positives is available, the
information on the disease status of any particular agent in the
population is not known.  

In~\cite{kim2020feedback}, the FPF algorithm is extended for a model 
with $M$ agents and $M$ 
observations.  The $M$ observations are non-agent specific.  Therefore, in its basic form, the problem is
characterized by data association uncertainty whereby the association
between the observations and agents must be deduced in addition to 
the agent state.  In~\cite{kim2020feedback}, the large-$M$ limit is
interpreted as a problem of collective inference.  This viewpoint is used to derive the
equation for the empirical distribution of the hidden agent states.
An FPF algorithm for this problem is presented and
illustrated via numerical simulations.  Formulae are described for
both the
Euclidean and the finite state-space case.  The classical FPF algorithm is shown to be the special case
(with $M=1$) 
of these more general results.  The simulations help show that
the algorithm well approximates
the empirical distribution of the hidden states for large $M$.   

Before closing this section, we remark on the Stratonovich form of the
mean-field FPF SDE~\eqref{eq:intro-FPF-Xbar}.  The FPF
is expressed in this form because of two reasons: 
\begin{enumerate}
\item The feedback control law is ``gain times error''
  which is appealing to control engineers, and structurally similar to
  the update formula in a Kalman filter.  Moreover, for the linear
  Gaussian model, the gain is the Kalman gain. 
\item Expressed in its Stratonovich form, the gain times error 
formula carries over to the Riemannian manifolds settings.  This is
because of the intrinsic nature of the Stratonovich form~\citep[Remark
1]{zhang2017feedback}.  
\end{enumerate}   
Notably, for the linear Gaussian model, the gain function is a constant
(i.e., does not depend upon $x$) and therefore the Stratonovich form and the  It\^o form are the
same.  For the general case, the It\^o form involves a Wong-Zakai
correction term as described in the following remark.

\begin{remark}[It\^o form of FPF]
	In its It\^o form, the mean-field
        FPF~\eqref{eq:intro-FPF-Xbar} is expressed as
	\begin{align*}
		\ud \bar{X}_t = &a(\bar{X}_t) \ud t + \sigma(\bar{X}_t)\ud \bar{B}_t+ \K_t(\bar{X}_t)  (\ud Z_t -
			\frac{h(\bar{X}_t) + \bar{h}_t}{2}\ud t) \\&+ \frac{1}{4}\sum_{j=1}^\Zdim \nabla | \K^{(j)}_t(\Xbar_t)|^2\ud t,
	\end{align*}  
where  $ \frac{1}{4}\sum_{j=1}^\Zdim \nabla | \K^{(j)}_t(\Xbar_t)|^2$ is the Wong-Zakai
correction term. The It\^o-Stratonovich relationship discussed here is based on interpreting  	$\mathsf K_t(x)$ as a function of space $x$ and time $t$, and interpreting  the
$\circ$ in the Stratonovich form only with
respect to the space $x$. In a recent paper~\cite[Sec. 3]{pathiraja2020mckean}, the gain
	function is defined and interpreted as a function of space $x$ and
	the density.  This is natural because the dependence upon
	time $t$ comes because of the changes in density ($\bar p_t$) as
	the time evolves.  Because the density is a stochastic process, it is
	argued that the appropriate interpretation of $\circ$ in the
	Stratonovich form should involve {\em both} space $x$ and the density.
	Using such an interpretation, the Stratonovich form involves extra-terms that are solutions to accompanying Poisson equations. 
\end{remark}

\section{Algorithms for gain function approximation}
\label{sec:gain-approx}
The exact gain $\K$ is a $d\times \Zdim$ matrix-valued function, where the $j$-th column of $\K$ is the solution of the  Poisson equation~\eqref{eq:intro-Poisson} for $j=1,\ldots,\Zdim$.  For the ease of presentation, the exposition in this section  is restricted to the scalar-valued observation setting, i.e. $\Zdim=1$, so that $\K$ becomes a $d$-dimensional vector-valued function and  the  superscript $j$ is dropped from the Poisson equation~\eqref{eq:intro-Poisson}.

\begin{figure}[t]
	\centering
\includegraphics[width=0.8\hsize]{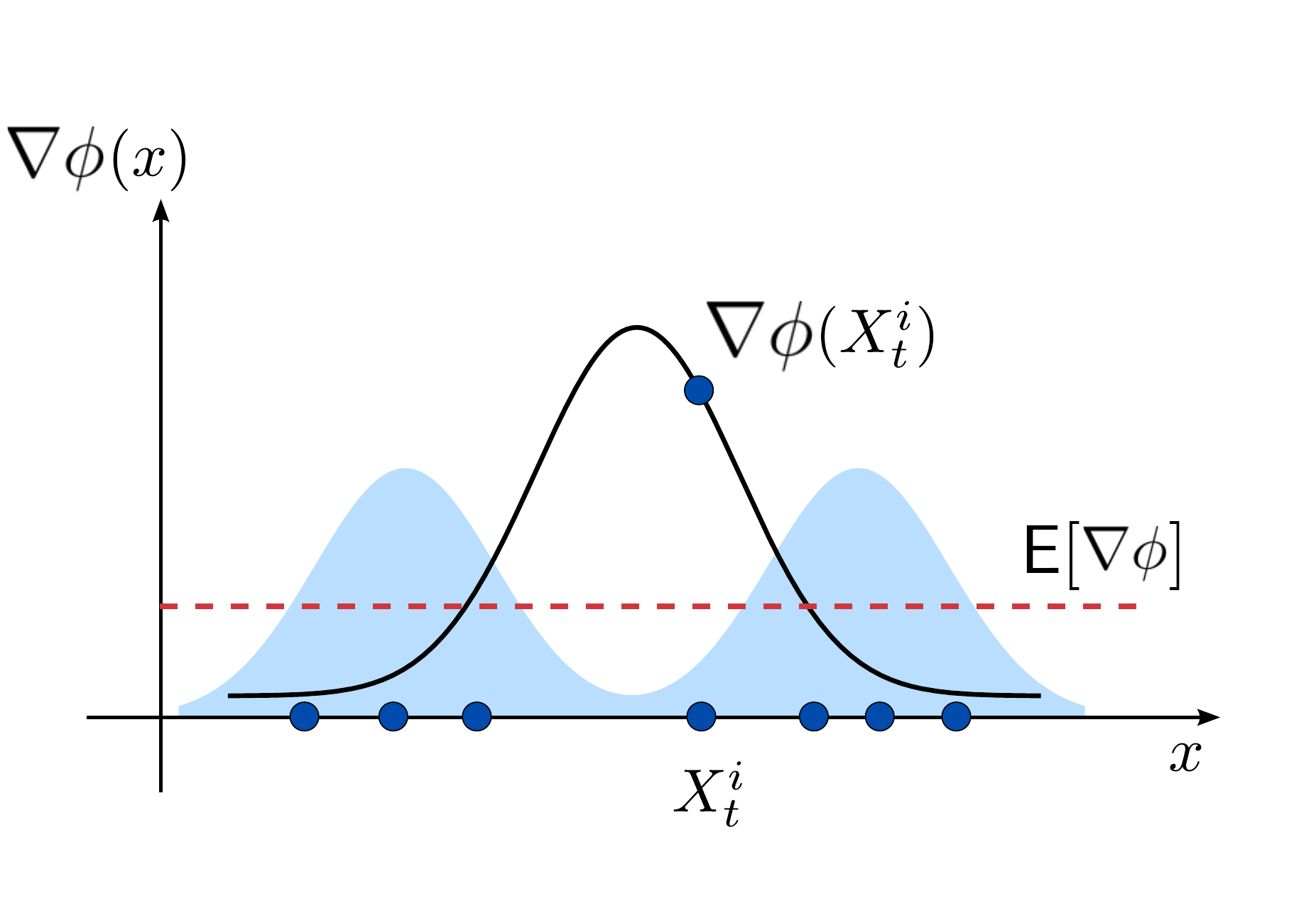}
	\caption{Gain function approximation problem in the feedback particle filter. The exact gain function
		$\K(x)= \nabla \phi(x)$ where $\phi$ solves the Poisson
		equation~\eqref{eq:intro-Poisson}. The numerical problem is
		to approximate $\K^i=\left. \nabla
		\phi(x)\right|_{x=X^i}$ using only the particles $\{X^i:
		1\leq i\leq N\}$ sampled from density $\rho$ (depicted as
		shaded region). The dashed line indicates the constant gain approximation, where the gain function is approximated by its expected value according to~\eqref{eq:const-gain-approx}.}
	\label{fig:gain-function-approx}
\end{figure}

In practice, the Poisson equation must be solved numerically.  The
numerical gain
function approximation problem is as follows:
\begin{align*}
	\text{input:} \quad& \text{samples $\{X^i:1\leq i\leq N\} \overset{\text{i.i.d.}}{\sim} \rho $,  $h(\cdot)$}\\
	\text{output:}\quad& \text{gain function $\{\K^i: 1\leq i \leq N\}$}
\end{align*}
where $\rho$ is the (posterior) density and $\K^i := \K(X^i)$.  The
explicit dependence on time $t$ is suppressed in this section. An
illustration of the gain
function approximation problem appears in
Fig.~\ref{fig:gain-function-approx}.  

\subsection{Motivation and overview of approaches}
\label{sec:sidebar}
The Poisson equation is a linear PDE.  In order to motivate the
various solution approaches, it is useful to first consider a
finite-dimensional counterpart
\begin{equation}
	Ax = b,
	\label{eq:PE_finite_dim}
\end{equation}
where $A$ is a $n\times n$ (strictly) positive-definite symmetric
matrix and the righthand-side $b$ is a given $n\times 1$ vector.  The
problem is to compute the unknown $n\times 1$ vector $x$.  For this purpose, the
following equivalent formulations of the finite-dimensional problem are first
introduced: 
\begin{enumerate}
	\item $x$ is the solution of the weak form
	\begin{equation*}
		y^\transpose A x = y^\transpose b,\quad \forall \; y \in \Re^n.
	\end{equation*}
	\item  For some chosen positive $\epsilon$, $x$ is the solution to the fixed-point equation
	\begin{equation*}
		x = e^{-\epsilon A} x + \int_0^\epsilon e^{-sA} b \; \ud s.
	\end{equation*}
	\item $x$ is the solution of an optimization problem
	\begin{equation*}
		x = \argmin_{z\in \Re^n}~\half z^\transpose A z - z^\transpose b.
	\end{equation*}
\end{enumerate}
When $n$ is large, these formulations are useful to numerically approximate the
solution of~\eqref{eq:PE_finite_dim}:
\begin{enumerate}
	\item For each fixed $y\in\Re^n$, the weak form is a single equation.
	By restricting $y$ to a suitable low-dimensional subspace 
	$S\subset\Re^n$, the number of linear equations is reduced for the
	purposes of obtaining an approximate solution (possibly also in $S$).  
	\item The fixed-point equation is useful because
          $e^{-\epsilon A}$ is a strict 
	contraction for $\epsilon>0$ (because $A$ is strictly positive-definite).  So, a good initial guess for
	$x$ can readily be improved by using the Banach iteration.  
	\item The optimization form is useful to develop alternate (e.g.,
	search type) algorithms to obtain the solution. 
\end{enumerate}

With this background, we turn our attention to the Poisson
equation~\eqref{eq:intro-Poisson} expressed succinctly as 
\begin{equation*}
	-\Delta_\rho \phi = (h-\bar{h}),
\end{equation*}
where $\bar{h}:=\int h(x) \rho(x)\ud x$ and
$\Delta_\rho:=\frac{1}{\rho}\nabla \cdot (\rho \nabla)$.  The linear operator
$\Delta_\rho$ is
referred to as the probability weighted Laplacian.  
Functional analytic considerations require introduction of the
function spaces: $L^2(\rho)$ is the space of square integrable
functions with respect to $\rho$ with inner product $ \langle
f,g\rangle := \int f(x)g(x)\rho(x)\ud x$; $H^1(\rho)$ is the
Hilbert space of functions in $L^2(\rho)$ whose first derivative, defined
in the weak sense, is the also in $L^2(\rho)$; and
$H^1_0(\rho)=\{\psi\in H^1(\rho) | \int \psi(x) \rho (x) \ud x = 0\}$.

These definitions are important because $H^1_0(\rho)$ is the natural
space for the solution $\phi$ of the Poisson
equation~\eqref{eq:intro-Poisson}.  The operator $-\Delta_\rho$ is
symmetric (self-adjoint) and positive definite because 
\begin{equation*}
	- \langle f,\Delta_\rho g \rangle= 
	\langle \nabla f,\nabla g\rangle
	= - \langle \Delta_\rho f,g \rangle,\quad \forall f,g\in H_0^1(\rho).
\end{equation*}
In the infinite-dimensional settings, one requires an additional technical condition---the 
Poincar\'e inequality (PI)---to conclude that the operator is in fact 
strictly positive-definite~\citep[Sec. 2.2]{taghvaei2020diffusion}.  Assuming the PI holds,
it is also readily shown that $\Delta_\rho^{-1}$ is well defined, i.e.,
a unique solution $\phi\in H^1_0(\rho)$ exists for any given $h\in
L^2(\rho)$~\cite[Thm. 2]{yang2016}.

For the purposes of numerical approximation, entirely analogous to the
finite-dimensional case, the following equivalent formulations of the Poisson
equation are introduced:
\begin{enumerate}
	\item $\phi$ is a solution of the weak form
	\begin{equation}
		\langle \nabla \psi, \nabla \phi \rangle= \langle \psi,h-\bar{h}\rangle  \quad \forall\; \psi
		\in H_0^1(\rho).\label{eq:Poisson-weak}
	\end{equation}
	\item For some chosen positive $\epsilon$, $\phi$ is a solution of the fixed-point equation
	\begin{equation}
		\phi = e^{\epsilon \Delta_\rho} \phi + \int_0^\epsilon
                e^{s\Delta_\rho} (h-\bar{h})\ud s. \label{eq:Poisson-semigroup}
	\end{equation}
The notation $e^{\epsilon \Delta_\rho}$ is used to denote the
semigroup associated with $\Delta_\rho$~\citep{bakry2013}.  The
semigroup is readily shown to be a Markov operator.  
	\item $\phi$ is the solution of an optimization problem
	\begin{equation}
		\phi = \argmin_{f \in H^1_0(\rho)} \;\; \half \langle \nabla f, \nabla
		f \rangle+ \langle f,h-\bar{h}\rangle.
		\label{eq:Poisson-opt}
	\end{equation}
\end{enumerate}

Each of the three formulations has been used to develop numerical
algorithms for gain function approximation.  A review of the resulting
constructions appears in the following three subsections:

\subsection{Galerkin and constant gain approximation}
\label{sec:gal_approx}
The starting point is the weak
form~\eqref{eq:Poisson-weak}. A relaxation is considered whereby $\psi
\in S = \text{span}\{\psi_1,\ldots,\psi_M\}$, a finite-dimensional
subspace of $H_0^1(\rho)$.  The functions $\psi_1,\ldots,\psi_M$ need
to be picked and are
referred to as the basis functions.  The
resulting algorithm is referred to as
the Galerkin algorithm~\citep[Sec 3.3]{yang2016}.  The algorithm is
given in Table~\ref{algo:gain_Galerkin}.

\begin{algorithm}
\label{algo:gain_Galerkin}
\caption{Synthesis of the gain function: Galerkin approximation}
\begin{algorithmic}[1]
	\REQUIRE $\{X^i\}_{i=1}^N$, $\{h(X^i)\}_{i=1}^N$, basis functions $\{\psi_l(x)\}_{l=1}^L$.
\ENSURE $\{\K^i\}_{i=1}^N$. \medskip
\STATE Calculate ${h}^{(N)}=\frac{1}{N}\sum_{i=1}^N h(X_t^i)$.
\STATE Calculate $b_k = \frac{1}{N} \sum_{i=1}^N (h(X_t^i) - {h}^{(N)}) \psi_k (X_t^i)$.
\STATE Calculate $A_{kl} = \frac{1}{N} \sum_{i=1}^N \nabla \psi_l(X_t^i)^\transpose \nabla \psi_k (X_t^i)$.
\STATE Solve the linear matrix equation $A \kappa = b$ for $\kappa$, where $A=[A_{kl}]$ and $b=[b_k]$.
\STATE$\K^i = \sum_{l=1}^L \kappa_l \nabla \psi_l(X_t^i)$.
\end{algorithmic}
\end{algorithm}

\begin{algorithm}
\label{algo:gain_cG}
\caption{Synthesis of the gain function: constant gain approximation}
\begin{algorithmic}[1]
			\REQUIRE $\{X^i\}_{i=1}^N$, $\{h(X^i)\}_{i=1}^N$. 
	\ENSURE $\{\K^i\}_{i=1}^N$. \medskip
\STATE Calculate $\hat{h}^{(N)}=\frac{1}{N}\sum_{i=1}^N h(X_t^i)$.
\STATE $\K^i = \frac{1}{N} \sum_{j=1}^N X_t^j \left(h(X_t^j) - \hat{h}^{(N)}\right)$
\end{algorithmic}
\end{algorithm}

The most important special case of the Galerkin algorithm is obtained upon picking
$S$ to be the subspace spanned by the $d$ coordinate functions
$\{x_1,x_2,\hdots,x_d\}$.  The special case yields the {\em constant gain
	approximation} of the gain $\K$ as
      its expected value.  
Remarkably, the
expected value admits a closed-form expression 
which is then readily approximated empirically using the particles:
\begin{equation}
	\begin{aligned}
		\K^{(\text{cnst. apprx.})} := \int \nabla \phi(x)
                \rho(x) & \ud x  = \int (h(x)-\bar{h})
		x \rho(x) \ud x \\&\approx \frac{1}{N}\sum_{i=1}^N\; (h(X^i)-{h}^{(N)}) X^i,
	\end{aligned}\label{eq:const-gain-approx}
\end{equation}
where ${h}^{(N)}:=N^{-1} \sum_i h(X^i)$.  
(See~\Fig{fig:gain-function-approx} for an illustration of the constant
gain approximation.) 
With the constant gain approximation, the FPF algorithm is a nonlinear
EnKF algorithm~\citep{TaghvaeiASME2017}. While its derivation starting
from an FPF 
is novel, the formula~\eqref{eq:const-gain-approx} has been used as a
heuristic in the
EnKF literature~\citep{evensen2006,bergemann2012ensemble}.

The main issue with the Galerkin approximation is that it is in
general very
difficult to pick the basis functions.  There have been a number of
studies to refine and improve upon this
formula~\citep{yang2016,yang2013,berntorp2016,matsuura2016suboptimal,Sean_CDC2016,radhakrishnan2018feedback,berntorp2018comparison}. 
In the following two subsections, we describe two approximations which
appear to be more promising approaches in general settings.

\subsection{Diffusion map-based algorithm}
\label{sec:diff_map}

The starting point is the fixed-point
equation~\eqref{eq:Poisson-semigroup} based on the Markov semigroup
$e^{\epsilon\Delta_{\rho}}$.  For small values of $\epsilon$,
there is a well known approximation of 
$e^{\epsilon\Delta_{\rho}}$ in terms of the so-called diffusion map
(which too is a Markov operator):  
\begin{equation}\label{eq:Teps-definition}
	(\Teps f)(x) := \frac{1}{n_\epsilon(x)}\int_{\Re^d}\frac{g_\epsilon(|x-y|)}{\sqrt{\int g_\epsilon(|y-z|)\rho(z)\ud z}}f(y)\rho(y)\ud y,
\end{equation} 
where $g_\epsilon(z):=e^{-\frac{z^2}{4\epsilon}}$ is the Gaussian
kernel in $\Re$ and $n_\epsilon(x)$ is the normalization factor chosen
so that $\int (\Teps 1)(x) \ud x = 1$~\citep{coifman}.  A
representative  approximation result is as follows:

\begin{proposition}[Prop. 3.4 in \citep{taghvaei2020diffusion}] \label{prop:Tepsn-convergence}
        Let $n \in \mathbb{N}$, $t_0<\infty$, and $t \in
	(0,t_0)$ with $\epsilon = \frac{t}{n}$. Then,
	for all functions $f$ such that $f,\nabla f \in L^4(\pr)$:
		\begin{equation*}
			\|(T_{\frac{t}{n}}^n - e^{t\Delta_{\rho}})f\|_{L^2(\pr)} \leq \frac{{t^\frac{3}{2}}}{n}C(\|f\|_{L^4(\pr)}+\|\nabla f\|_{L^4(\pr)}),
		\end{equation*}
		where the constant $C$ depends only on $t_0$ and $\pr$.
\end{proposition}

Because the diffusion map~\eqref{eq:Teps-definition} is defined using
Gaussian kernels, its empirical approximation is straightforward:
\begin{equation*}
	(\TepsN f)(x) =  \frac{1}{n_\epsilon^{(N)}(x)}\sum_{i=1}^N \frac{g_\epsilon(|x-X^i|)}{\sqrt{\sum_{j=1}^N g_\epsilon(|X^i-X^j|)}}f(X^i),
\end{equation*} 
where $n_\epsilon^{(N)}(x)$ is the normalization factor.  The
nature of the approximation is as follows:

\begin{proposition}[Prop. 3.5 in \cite{taghvaei2020diffusion}] \label{prop:TepsN-convergence}
	Consider the diffusion map kernel $\Teps$ and its empirical approximation $\{\TepsN\}_{N\in \mathbb{N}}$. Then for any bounded continuous function $f \in C_b(\Re^d)$:
	\begin{enumerate}  
			\item (Almost sure convergence) For all $x \in \Re^d$
			\begin{equation*}
					\lim_{N \to \infty} (\TepsN f)(x) = (\Teps f)(x),\quad \text{a.s.}
				\end{equation*}
			\item (Convergence rate) For any $\delta \in (0,1)$, in the asymptotic limit as $N \to \infty$,  
			\begin{equation*}
					\int |(\TepsN f)(x) - (\Teps f)(x)|^2\pr(x)\ud x \leq O(\frac{\log(\frac{N}{\delta})}{N\epsilon^{d}}),
				\end{equation*}
			with probability higher than $1-\delta$.
		\end{enumerate}
\end{proposition}

\begin{figure*}[t]
	\centering
	\begin{tabular}{cc}

			\subfigure[]{\includegraphics[width=0.9\columnwidth]{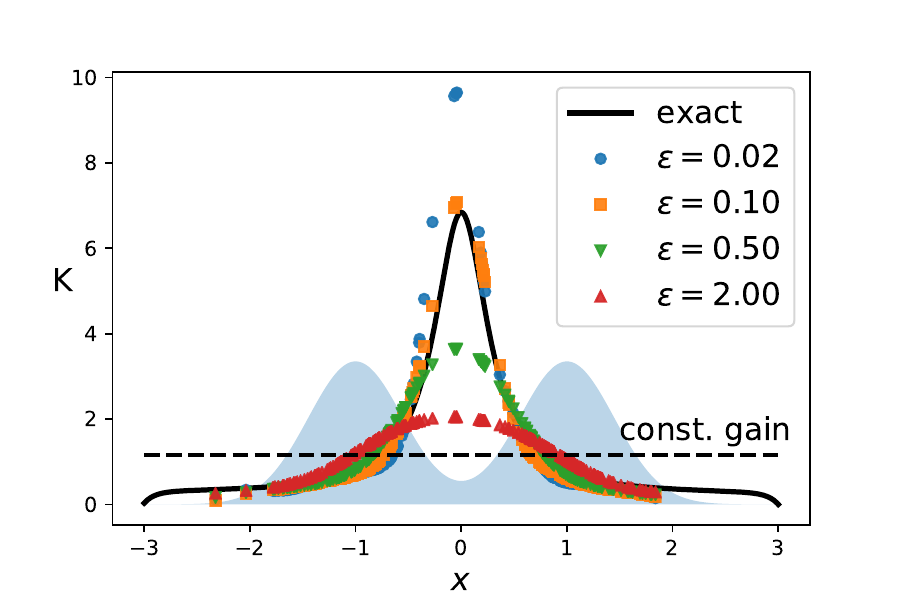}}

		& 
			\subfigure[]{\includegraphics[width=0.9\columnwidth]{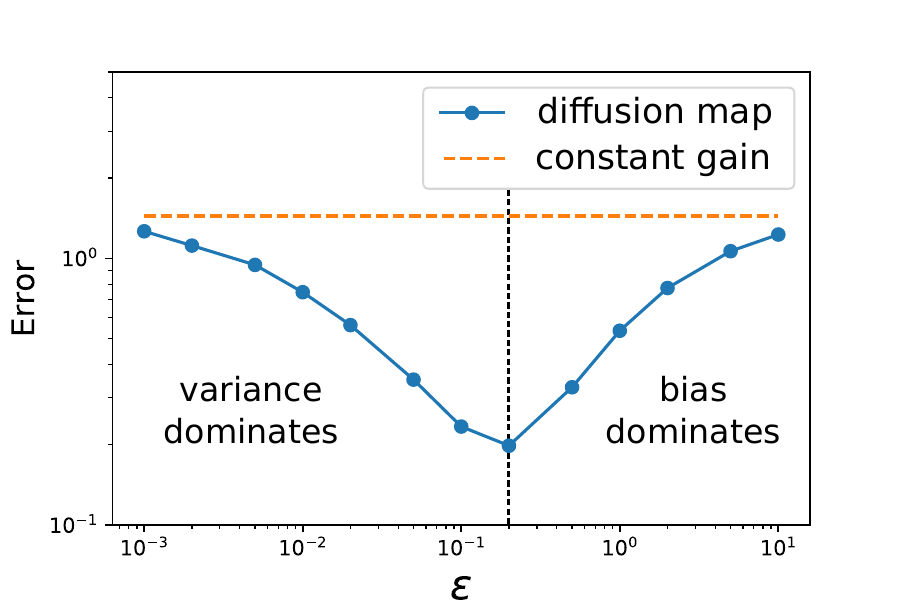}}
		
	\end{tabular}
	\caption{Bias variance trade-off
		in the diffusion map-based gain function approximation
                algorithm:
                (a) Gain function computed for different values of
                $\epsilon$ with $N=200$ particles.  The dashed line is the constant
		gain solution~\eqref{eq:const-gain-approx}. As
		$\epsilon$ gets larger, the diffusion map gain converges
		to the constant gain.  (b)  Plot of the 
                MSE as a function of $\epsilon$. The shaded area in the background 
		of part~(a) is the density $\rho$ which is taken as sum of two Gaussians
		${\cal N}(-1,\sigma^2)$ and ${\cal N}(+1,\sigma^2)$
		with $\sigma^2=0.2$.  The exact gain function $\K(x)$
		is computed for $h(x)=x$ by using an (exact) integral
		formula forr the solution~\cite[Eq. 4.6]{taghvaei2020diffusion}. 
		In part~(b), the MSE is computed as an empirical
		approximation of the lefthand-side of~\eqref{eq:MC_error} by averaging
		over $1000$ simulation runs. }
	\label{fig:kernel-approx}
\end{figure*} 

With these approximations, the fixed-point
equation~\eqref{eq:Poisson-semigroup} is approximated in two steps: 
\begin{subequations}
\begin{enumerate}
\item The semigroup $e^{\epsilon\Delta_{\rho}}$ is approximated by
the diffusion map $\Teps$:
\begin{align} \label{eq:phieps}
&\text{(step 1)}\qquad	\phieps = \Teps \phieps  + \epsilon (h -
                                  \bar{h}_\epsilon),
\end{align}
where $\bar{h}_\epsilon =  \int h(x)\rhoeps(x)\ud x$ with $\rhoeps(x)
= \frac{\neps(x)\rho(x)}{\int \neps(x)\rho(x)\ud x}$.
\item $\Teps$ is approximated by
its empirical approximation $\TepsN$:
\begin{align}
&\text{(step 2)}\qquad	\phiepsN = \TepsN \phiepsN + \epsilon (h - \bar{h}_\epsilon^{(N)}), \label{eq:phiepsN}
\end{align}
where $\bar{h}_\epsilon^{(N)} = \int h(x)\rhoepsN(x)\ud x$ with $\rhoepsN(x)
=  \frac{\sum_{i=1}^N \neps(x)\delta_{X^i}}{\sum_{i=1}^N \neps(X^i)}$.  
\end{enumerate}
\end{subequations}

Based
on the finite-dimensional fixed-point equation~\eqref{eq:phiepsN}, an algorithm for gain function approximation is
given in Table~\ref{tab:DM-algorithm}. {In the algorithm, the gain function is approximated by the formula 
\begin{align*}
	\K_\epsilon^{(N)} = \nabla \left[  \TepsN \phiepsN  + \epsilon\TepsN (h-\bar h_\epsilon^{(N)}) \right]. 
\end{align*}
There are alternative ways to approximate the gain function in terms of $\phiepsN$.  While these solutions have the same asymptotic in the limit as $\epsilon \to 0$, they behave differently when $\epsilon$ is large. The specific approximation selected here does not require derivative of the observation function and converges to the constant gain approximation as $\epsilon$ becomes large~\cite[Remark 4.8]{taghvaei2020diffusion}.}

\begin{algorithm}[t]
		\caption{Synthesis of the gain function: diffusion map-based algorithm}
	\begin{algorithmic}[1]
		\REQUIRE $\{X^i\}_{i=1}^N$, $\{h(X^i)\}_{i=1}^N$, $\phivec_{\text{prev}}$, $\epsilon$, L.
		\ENSURE $\{\K^i\}_{i=1}^N$. \medskip
		\STATE Calculate $g_{ij}:=e^{-\frac{|X^i-X^j|^2}{4\epsilon}}$ for $i,j=1$ to $N$.\medskip
		\STATE Calculate $k_{ij}:=\frac{g_{ij}}{\sqrt{\sum_l g_{il}}\sqrt{\sum_l g_{jl}}}$ for $i,j=1$ to $N$.
		\STATE Calculate $d_i= \sum_{j} k_{ij}$ for $i=1$ to $N$.
		\STATE Calculate $\Ten_{ij}:=\frac{k_{ij}}{d_i}$ for $i,j=1$ to $N$.
		\STATE Calculate $\pi_i=\frac{d_i}{\sum_{j}d_j}$ for $i=1$ to $N$.
		\STATE Calculate $\hat{\hvec}= \sum_{i=1}^N \pi_jh(X^i)$. \medskip
		\STATE Initialize $\phivec=\phivec_{\text{prev}}$.
		\medskip
		\FOR {$t=1$ to  L}
		\STATE  $\phivec_i= \sum_{j=1}^N \Ten_{ij} \phivec_j + \epsilon (\hvec-\hat{\hvec})$ for $i=1$ to $N$.\medskip
		\ENDFOR
		\STATE Calculate $r_i = \phivec_i + \epsilon \hvec_i$ for $i=1$ to $N$.
		\STATE Calculate $s_{ij} = \frac{1}{2\epsilon}\Ten_{ij}(r_j-\sum_{k=1}^N
		\Ten_{ik}r_k)$ for $i,j=1$ to $N$.
		\STATE Calculate $\K^i = \sum_j s_{ij}X^j$ for $i=1$ to $N$.
	\end{algorithmic}
	\label{tab:DM-algorithm}
\end{algorithm}

\begin{figure*}[t]
	\centering
	\begin{tabular}{cc}
		\subfigure[]{
			\includegraphics[width=0.9\columnwidth]{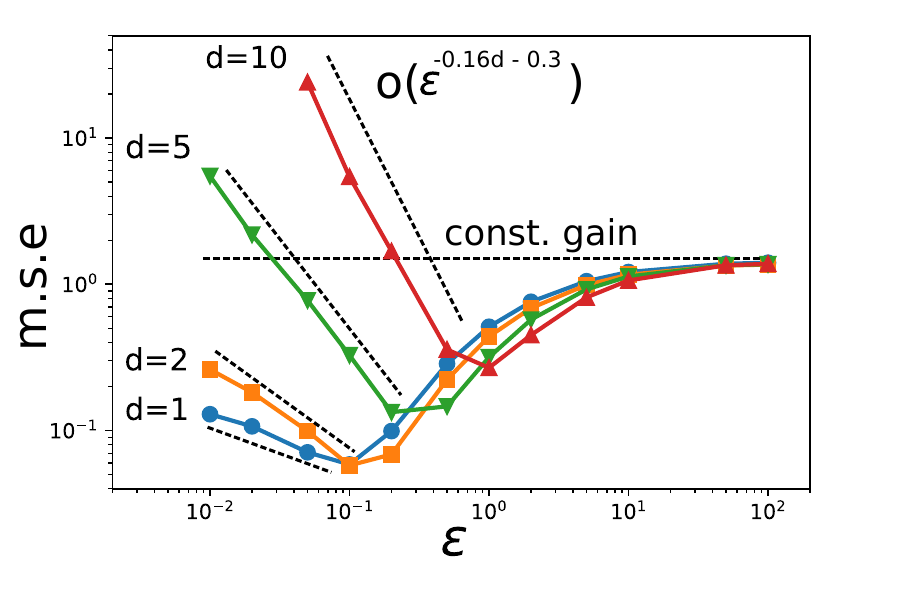}
		}&
		\subfigure[]{
			\includegraphics[width=0.9\columnwidth]{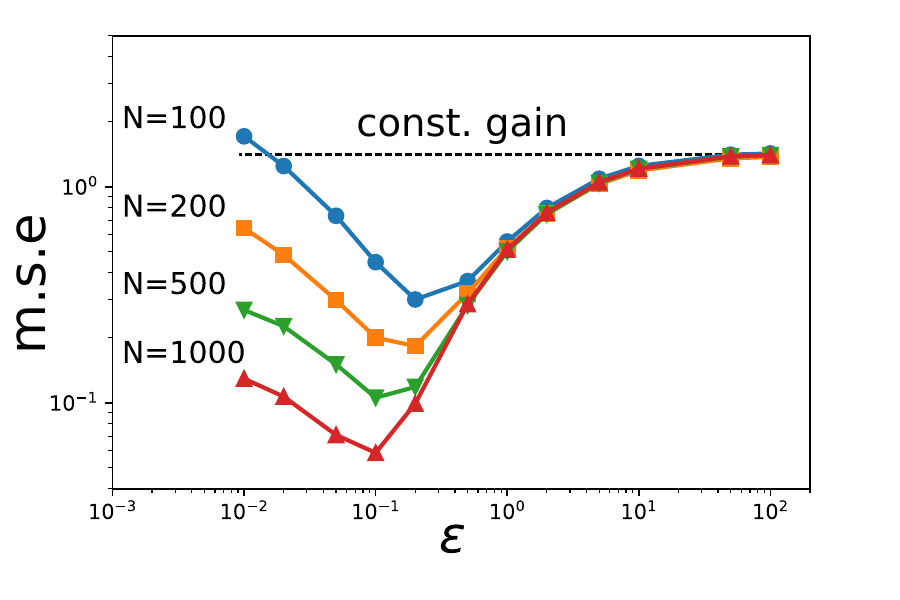}
		}		
	\end{tabular}
	\caption{Bias-variance trade-off as a function of (a) the state
          dimension $d\in\{1,2,5,10\}$ (for a fixed $N=1000$); and (b)
          the number of particles $N\in\{100,200,500,1000\}$ (for a fixed
          $d=1$).  In the vector case, $\pr(x) = \rho_{b}(x_1)
          \prod_{n=2}^d\rho_g(x_n)$ where $\rho_b$ is the bimodal
          density (same as in \Fig{fig:kernel-approx}) and
          $\rho_g$ is the Gaussian density.  
	}
	\label{fig:numerics-all}
\end{figure*}

\paragraph{Error analysis}
The error in diffusion map approximation comes from two sources: 
\begin{enumerate}
\item The bias error due to the diffusion map approximation of the
semigroup (step 1); and
\item The variance error due to empirical approximation in
terms of particles (step 2).
\end{enumerate} 
The error is analyzed
in~\citep{taghvaei2020diffusion} where the following result is proved:
\begin{proposition}[Thm. 4.3 and 4.4 in~\citep{taghvaei2020diffusion} ]\label{prop:DM-error}
	Consider the fixed-point formulation of the Poisson equation~\eqref{eq:Poisson-semigroup}, its diffusion-map approximation~\eqref{eq:phieps}, and its empirical approximation~\eqref{eq:phiepsN}. 
		\begin{enumerate}
		\item For each fixed $\epsilon>0$, there exists a unique solution to~\eqref{eq:phieps} with a uniform bound $	\|\phieps\|_{L^2(\preps)} \leq C\|h\|_{L^2(\preps)}$.  In the asymptotic limit as $\epsilon \to 0$
		\begin{equation*}
			\|\phieps- \phi\|_{L^2(\preps)} \leq O(\epsilon).
		\end{equation*} 
	\item The operator $\TepsN$  is a strict contraction  on  $L^2_0(\rhoepsN)$ and the fixed-point equation~\eqref{eq:phiepsN} admits a unique solution.  The approximate solution
	$\phiepsN$ converges to the kernel solution $\phieps$   
	\begin{equation*}
		\lim_{N \to \infty} \|\phiepsN -\phieps\|_{L^\infty(\Omega)} = 0,\quad \text{a.s.}
	\end{equation*}
	\end{enumerate}
\end{proposition}

The following diagram illustrates the convergence and the respective
types of errors:
\begin{equation*}
	\phiepsN \underset{\text{(variance)}}{\overset{N \uparrow \infty}{\longrightarrow}} \phieps
	\underset{\text{(bias)}}{\overset{\epsilon \downarrow 0}{\longrightarrow}} \phi. 
\end{equation*}
A quantitative bound on the mean-squared error (MSE) is obtained in the asymptotic limit as $\epsilon\downarrow
0$ and $N\to\infty$ as follows:
\begin{align}
\underbrace{\left(\mathbb{E}[ \frac{1}{N}\sum_{i=1}^N | {\sf
		K}^i - \nabla\phi (X^i) |^2]\right)}_{\text{MSE}} \leq
  \underbrace{O(\epsilon^2)}_{\text{bias}} +
  \underbrace{O(\frac{1}{\epsilon^{(2+d)} N})}_{\text{variance}},
	\label{eq:MC_error}
\end{align}
where $\{{\sf K}^i\}_{i=1}^N$ is computed from the Algorithm
(Table~\ref{tab:DM-algorithm}) and $\nabla\phi$ is the exact gain function
from solving the Poisson equation~\eqref{eq:intro-Poisson}. 
The error due to bias converges to zero as $\epsilon \to 0$ and the
error due to variance converges to zero as $N \to \infty$.  There is
trade-off between the two errors: To reduce bias, one must reduce
$\epsilon$.  However, for any fixed value of $N$, one can reduce
$\epsilon$ only
up to a point where the variance starts increasing.  
The bais-variance trade-off is illustrated with the aid of a scalar
($d=1$) example in 
Fig.~\ref{fig:kernel-approx}:  If $\epsilon$ is large, the error
due to bias dominates, while if $\epsilon$ is small, the error due to
variance dominates.  An numerical illustration of scalings with $N$
and $d$ appears in Fig.~\ref{fig:numerics-all}.  Additional details on
both these examples can be found in~\cite[Sec.~5]{taghvaei2020diffusion}.

\begin {table*}[h]
\caption {Applications and evaluation of the feedback particle filter} 
\label{tab:applns} 
\begin{center}
 \begin{tabular}{| l | l | l | l |} 
 \hline
 Authors & Applications of FPF & Reference & Year \\ [0.5ex] 
 \hline\hline
 del Moral and Horton & Quantum harmonic oscillators &
                                                  \cite{del2021quantum}
                                      & 2021 \\ \hline 
Wang et. al. & Unmanned aerial vehicle tracking &
                                                  \cite{wang2021quantized}
                                      & 2021 \\ \hline
Su et. al. & Soil estimation & \cite{su2019online} & 2021 \\ \hline
Kumar and Mishra & Marine applications & \cite{zheng2019parameter} &
                                                                     2019 \\ \hline
Berntorp and Grover & Satellite tracking and re-entry &
 \cite{berntorp2015} & 2015 \\ \hline
 Surace et. al. & Evaluation and comparison of FPF &
                                                           \cite{surace_SIAM_Review} & 2017 \\ \hline
Stano & Hopper-dredger model  & \cite{stano2018collapse,stano2014} &
                                                                      2014 \\ \hline
Matsuura et. al. & Target state estimation & \cite{matsuura2016suboptimal} & 2016 \\ \hline
Kutschireiter et. al. & Neuronal dynamics &
                                            \cite{kutschireiter2017nonlinear}
                                      & 2016 \\ \hline
Tilton et. al. & Coupled oscillators & \cite{Tilton_coupled_oscillator_FPF} & 2013
\\ \hline
Tilton et. al. & Marine estimation & \cite{adamtilton_fusion13} & 2013 \\
 \hline
\end{tabular}
\end{center}
\vspace{-0.3in}
\end{table*}

\begin{remark}[Relationship to the constant gain formula~\eqref{eq:const-gain-approx}]
There is a remarkable and somewhat unexpected 
relationship between the diffusion map and the constant gain
approximation~\cite[Prop. 4.7]{taghvaei2020diffusion}.  In particular, in the limit as
$\epsilon \to \infty$, the diffusion map gain converges to the
constant gain~\eqref{eq:const-gain-approx}.  This suggests a systematic procedure to improve 
upon the constant gain by de-tuning the value of
$\epsilon$ away from the [$\epsilon
= \infty$] limit.  For any fixed $N$, a finite value of $\epsilon$ is chosen to
minimize the MSE according to the bias variance trade-off.  Based
on this, a rule of thumb for choosing
the $\epsilon$ value appears in~\cite[Remark 5.1]{taghvaei2020diffusion}. 
\end{remark}

\begin{remark}[Analysis of FPF with diffusion map approximation] An
  analysis of the finite-$N$ FPF using the diffusion map approximation
  appears in~\citep{pathiraja2021analysis}.  Under mild technical
  conditions on the drift $a(\cdot),\sigma(\cdot),h(\cdot)$, it is
  shown that the finite-$N$ FPF is well-posed, i.e., a strong solution
  exists for all time
  $t$~\cite[Thm. 1.1]{pathiraja2021analysis}. Based on a propagation
  of chaos type analysis, convergence estimates are
  derived to relate the finite-$N$ system to
  its mean-field limit~\citep[Thm. 1.2]{pathiraja2021analysis}.  These
  estimates are shown to hold up to a certain stopping time.   For
  arbitrary time $t$, well-posedness and convergence remains an open problem.
\end{remark}

\subsection{Variational approximation}\label{sec:var-Poisson}

The starting point is the variational form~\eqref{eq:Poisson-opt}.
The objective function is denoted by $J(f)$ with its empirical
approximation is obtained as
\begin{equation*}
	J^{(N)}(f) := \frac{1}{N} \sum_{i=1}^N \half |\nabla f(X^i)|^2 - f(X^i)(h(X^i)-{h}^{(N)})
\end{equation*} 
The problem of minimizing the empirical approximation over all
functions is ill-posed: the minimum is unbounded and minimizer does
not exist.  (Abstractly, this is because the empirical probability distribution
does not satisfy the Poincar\'e inequality.)
Therefore, we consider  
\begin{equation*}
	\min_{f_\theta \in \calF_\Theta}~J^{(N)}(f_\theta)
\end{equation*}
where $\calF_\Theta$ is a parameterized class of functions. A function
in the class $\calF_\Theta$ is denoted by $f_\theta(x)$ or
$f(x;\theta)$ where $\theta \in \Theta$ is the parameter, and $\Theta$
is the parameter set. The two main examples are as follows:
\begin{enumerate}
	\item $\calF_\Theta = \{ \sum_{j=1}^M \theta_j \psi_j; ~
	\psi_j \in H^1_0, \theta_j \in \mathbb{R} \text{ for }
	j=1,\ldots,M\}$ is a linear combination of selected basis
	functions. With a linear parametrization, the solution of the
        empirical optimization problem is given by the Galerkin
        algorithm~\citep[Remark~5]{yang2016}. 
	\item $\calF_\Theta$ is a neural network where 
	the parameters $\theta$ are the weights in the network.   
\end{enumerate}

In practice, it is not possible to solve the optimization problem
exactly, but up to some optimization gap. In particular, let $\phi^{(N)}_\theta$ be the output of an optimization algorithm that solves the problem up to $\epsilon$ error, i.e.,
\begin{equation*}
	J(\phi^{(N)}_\theta)  \leq \min_{f \in H^1_0} J(f) + \epsilon.
\end{equation*}
 The good news is that it is
possible to upper-bound the error in approximating the gain function
in terms of this optimization gap. 

\begin{figure*}[t]
	\centering
	\includegraphics[width=0.95\columnwidth]{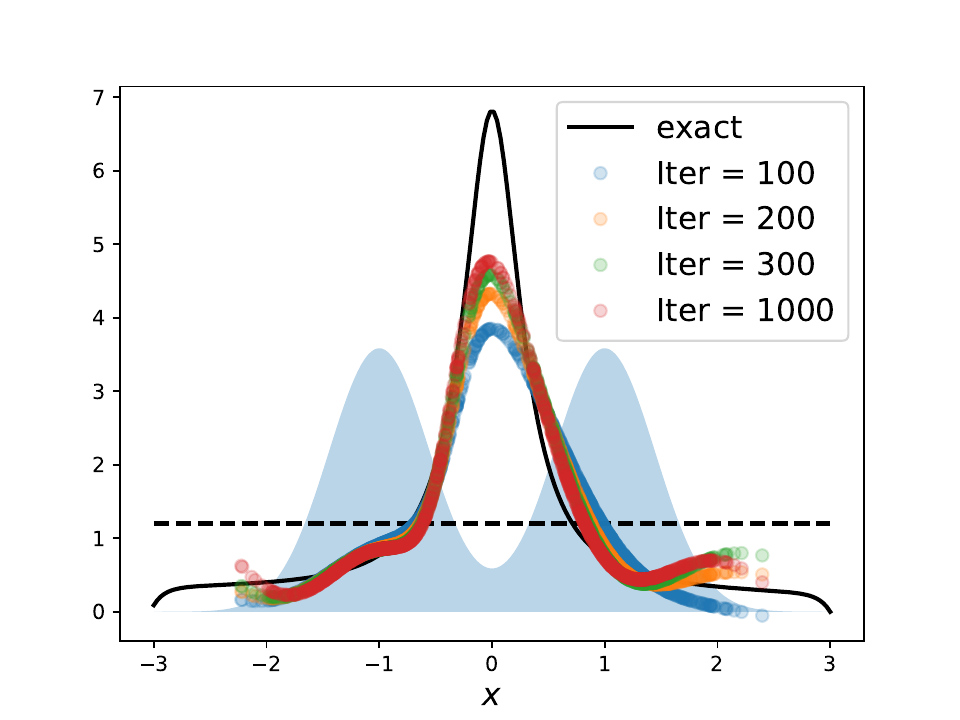} \hspace{10pt}
		\includegraphics[width=0.95\columnwidth]{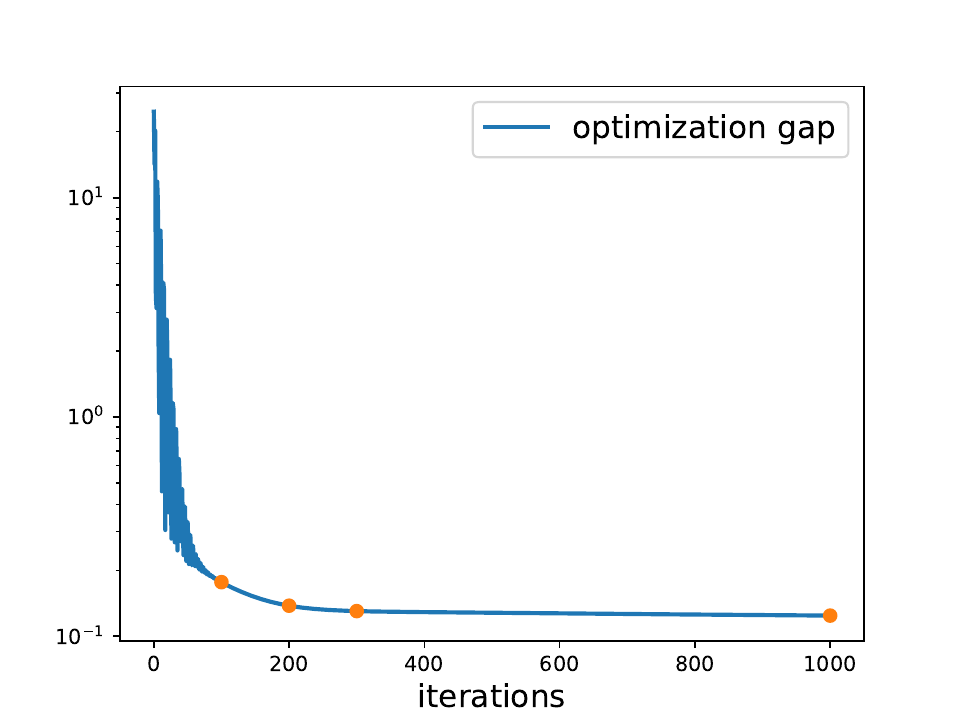}
		\caption{Results of the variational gain
                  function approximation using a neural network parameterization:  Plot of (a) the gain
                  function; and (b) the optimization gap as the number of
                  iterations of the Adam algorithm.  The problem setup is the same
                  as~\Fig{fig:kernel-approx}.  
}
		\label{fig:gain-approx-NN}
\end{figure*}

 \begin{proposition}[Prop. 1 in \cite{olmez2020deep}]\label{lem:J-stability}
Let $\K^{(N)}_\theta = \nabla \phi^{(N)}_\theta$ where $\phi^{(N)}_\theta$
is the output of an optimization algorithm that solves the
minimization objective $J(f)$ with $\epsilon$ optimality gap. Then
\begin{equation*}
	\|\K^{(N)}_\theta  -\K\|^2_{L^2_\rho} \leq  2\epsilon,
\end{equation*}
where $\K=\nabla\phi$ is the exact gain function. 
	 \end{proposition}
The optimization gap $\epsilon$ depends on the selected
parametrization $\mathcal F_\theta$, number of particles $N$, and the
iteration number of the employed optimization algorithm. Its
characterization and analysis is open and the subject of ongoing work. In general, such analysis falls under the framework of statistical learning theory~\citep{anthony1999neural,shalev2014understanding}. 

The numerical results using this approach are depicted
in~\Fig{fig:gain-approx-NN}.  These results are for the bimodal
example introduced in~\Fig{fig:kernel-approx}.   
The gain function is parameterized using
a two-layer residual NN with $32$ neurons per layer.  The
Adam algorithm is used to learn the parameters of the NN.  Additional
details on the numerics can be found in~\citep{olmez2020deep}.

{
\subsection{Numerical evaluation of FPF}

Numerical evaluations of the FPF algorithm, and comparisons with the
nonlinear extensions of the Kalman filer and conventional particle
filters, have been subject of several publications (some of these
studies are tabulated in the Table~\ref{tab:applns}).  Notable amongst
these is the early work of~\citep{berntorp2015} who both extended the
algorithm and applied it to two highly nonlinear applications
in aerospace, namely, the {\em re-entry} and {\em two-body} problems.
Another notable early work is~\citep{stano2014} on the application of
estimating soil-dependent time-varying parameters of the hopper
sedimentation model.  The study includes extensive comparisons with
the conventional particle filters. 
While these studies report favorable  accuracy  and computational
cost, the application of FPF to truly high-dimensional and nonlinear
problems remains still open.  In particular, beyond the toy examples,
we do not know of any application where the diffusion map
approximation has been applied.  

\begin{remark} Curse of dimensionality (CoD) is one of the main issues
   motivating the recent work on particle filters~\citep{,bengtsson08,bickel2008sharp,beskos2014error,rebeschini2015can}.  The analysis
   presented in \Sec{sec:FPF_comp} helps show that, at least in the
   linear Gaussian settings of the problem, the FPF/EnKF algorithm does not
   suffer from CoD.  This is because the Poisson equation admits an
   explicit solution in this case.  Because FPF is an exact algorithm, if/when CoD can be avoided in the
   nonlinear case really depends on the quality of the gain function
   approximation.  The bias-variance analysis of the diffusion map
   algorithm, presented in \Sec{sec:diff_map}, is helpful to see some
   of the tradeoffs.  The analysis suggests that to avoid CoD one must
   take advantage of (i) the underlying regularity of the
   gain function (e.g., constant in the linear Gaussian case), and/or
   (ii) 
   inherent low-dimensional structure in the problem (e.g.,
   approximation of posterior density in a low-dimensional
   manifold where a good diffusion map approximation can be
   obtained). One promising avenue is the variational gain function approximation using neural networks, as described
   in~\ref{sec:var-Poisson}. It remains to be seen whether some of the
   outstanding successes of neural networks in other fields can be
   replicated to avoid CoD in the particle filters.
\end{remark}
}
\section{Optimal transport theory}
\label{sec:OT-FPF}

In this section, we describe a systematic procedure to construct the
exact mean-field process $\bar{X}$ introduced as step 1
in~\eqref{eq:Xbar-u-K}.  The first aspect to note is that while the
FPF~\eqref{eq:intro-FPF-Xbar} provides an explicit formula for $u$ and $\K$, the formula is not unique: 
One can interpret~\eqref{eq:Xbar-u-K} as transporting
the prior density $p_0$ at time $t = 0$ to the posterior density $p_t$
at time t.  
Clearly, there are infinitely many maps that transport one
density into another. 
This suggests that there are infinitely many choices of control laws that all lead to exact filters.  
This is not surprising: The exactness condition specifies only the
marginal density at times $t$, which is not enough to uniquely identify a
stochastic process, e.g., the joint density at two time instants
has not been specified.

In the following, we first discuss the non-uniqueness issue for the
simpler linear Gaussian model.  The non-uniqueness naturally
motivates optimal transport ideas to uniquely solve for $u$ and
$\K$.  This is the subject of the remainder of this section to derive
the feedback control law for the FPF~\eqref{eq:intro-FPF-Xbar}.

\subsection{Non-uniqueness issue in linear-Gaussian setting} 
\label{sec:LG-OT}
Consider the linear Gaussian FPF~\eqref{eq:intro-FPF-Xbar-lin} for the
mean-field process~$\{\bar{X}_t\}_{t\geq 0}$.  The conditional mean and variance of $\bar{X}_t$ are denoted by
$\mbar_t$ and $\Sigmabar_t$, respectively.  The conditional mean
evolves according to 
\begin{equation*}
	\ud \mbar_t = A \mbar_t \ud t + \Kbar_t(\ud Z_t - H\mbar_t\ud t),
\end{equation*}
where $\Kbar_t:= \Sigmabar_t H^\transpose$.  
Define an error process $\xi_t := \Xbar_t -
\mbar_t$.  Its equation is given by
\begin{equation*}
	\ud \xi_t = (A - \frac{1}{2} \Sigmabar_t H^\transpose H)\xi_t + \sigma_B \ud \Bbar_t.
\end{equation*}
This is a linear system and therefore the variance of $\xi_t$, which
equals $\Sigmabar_t$ (by definition),
evolves according to the Lyapunov equation
\begin{align*}
	\frac{\ud}{\ud t} \Sigmabar_t &=  (A - \frac{1}{2} \Sigmabar_t
                                        H^\transpose H) \Sigmabar_t +
                                        \Sigmabar_t(A -
                                        \frac{1}{2}\Sigmabar_t
                                        H^\transpose H) ^\transpose + \Sigma_B \\&= \Ricc(\Sigmabar_t).
\end{align*}
The derivation helps show that the equations for the mean and variance
are identical to the Kalman filter equations,~\eqref{eq:intro-Kalman-mean}
and~\eqref{eq:intro-Kalman-cov}, respectively, and thus proves the exactness
property of the linear FPF~\eqref{eq:intro-FPF-Xbar-lin}.  

These arguments suggest the following general procedure to
construct an exact $\Xbar$ process:
Express $\Xbar_t$ as a sum of two terms: \[ \Xbar_t =\mbar_t +
  \xi_t,\quad t\geq 0,\] 
where $\mbar_t$ evolves according to~\eqref{eq:intro-Kalman-mean} and the evolution of $\xi_t$ is defined by the SDE:
\begin{equation*}
	\ud \xi_t = G_t \xi_t \ud t + \sigma_t \ud 
	\Bbar_t   +\sigma'_t \ud \Wbar_t,\label{eq:EG}
\end{equation*} 
where $\{\Wbar\}_{t\geq 0}$ and $\{\Bbar\}_{t\geq 0}$ are independent copies of
the measurement noise  $\{W\}_{t\geq 0}$ and the process noise
$\{B\}_{t\geq 0}$, respectively, and $G_t$, $\sigma_t$, and
$\sigma'_t$ satisfy the matrix equation (for each time)
\begin{equation}
	G_t \Sigmabar_t + \Sigmabar_t G_t^T + \sigma_t \sigma_t^\transpose
        + \sigma'_t( \sigma'_t)^\transpose= \Ricc(\Sigmabar_t),\quad t\geq 0.
	\label{eq:G-gen}
\end{equation}
By construction, the equation for the variance is given by the Riccati equation~\eqref{eq:intro-Kalman-cov}.
The result is summarized in the following Proposition:
\begin{proposition}[Prop. 1 in \cite{taghvaei2022optimality}]
	\label{prop:exactness-linear}
	Consider the linear-Gaussian filtering problem~\eqref{eq:model-linear} and the following family of the mean-field processes
	\begin{align*}
		\ud \bar X_t & = A \mbar_t \ud t + \Kbar_t(\ud Z_t - H\mbar_t\ud t)\\&+ G_t(\bar X_t - \bar m_t) \ud t  + \sigma_t \ud 
		\Bbar_t   +\sigma'_t \ud \Wbar_t,\; \bar{X}_0\sim N(m_0,\Sigma_0),
	\end{align*}
where $G_t$, $\sigma_t$, and $\sigma'_t$ satisfy the consistency
condition~\eqref{eq:G-gen}. Then, $\bar X_t$ is exact, i.e. the
density of $\bar X_t$ is Gaussian $N(\bar{m}_t,\bar{\Sigma}_t)$ where $\bar{m}_t$ and
$\bar{\Sigma}_t$ solve the Kalman filter
equations,~\eqref{eq:intro-Kalman-mean}
and~\eqref{eq:intro-Kalman-cov}, respectively. 
\end{proposition}

In general, with different choices of $\sigma_t$ and
$\sigma'_t$, there are infinitely many solutions for
\eqref{eq:G-gen}. Below, we describe three solutions that lead to
three established form of EnKF and linear FPF:
\begin{enumerate}
	\item EnKF with perturbed observation~{
		\cite[Eq. (27)]{reich11}}:
	\begin{align*}
		G_t = A - \Sigmabar_t H^\transpose H,\quad {  \sigma_t = \sigma_B
			,\quad \sigma'_t=\Sigmabar_t H^\transpose}.
	\end{align*}
	\item Stochastic linear FPF~\cite[Eq. (26)]{yang2016} or square-root form of the EnKF~\cite[Eq (3.3)]{bergemann2012ensemble} : 
	\begin{align*}
		G_t = A - \frac{1}{2}\Sigmabar_t H^\transpose H,\quad \sigma_t = 
		\sigma_B,\quad \sigma'_t=0.
	\end{align*}
	\item Deterministic linear FPF~\cite[Eq. (15)]{AmirACC2016}{ \cite[Eq. (82)]{jana2016stability}}: 
	\begin{align*}
		G_t = A - \frac{1}{2}\Sigmabar_t H^\transpose H + {\frac{1}{2}\Sigma_B\Sigmabar_t^{-1}},\quad \sigma_t = 0,\quad \sigma'_t=0.
	\end{align*}
\end{enumerate}

Fix $\sigma_t,\sigma'_t$.  Then given any particular solution $G_t$ of~\eqref{eq:G-gen}, one can construct a
family of solutions $G_t + \Sigmabar_t^{-1} \Omega_t$, where
$\Omega_t$ is any arbitrary skew-symmetric matrix \citep[Sec. III-B]{taghvaei2020optimal}. 
For the linear Gaussian problem, the non-uniqueness issue is well
known in literature:  The two forms of EnKF, the perturbed observation form~\citep{reich11} and the
square-root form~\citep{bergemann2012ensemble} are standard. 
A homotopy of exact deterministic and stochastic EnKFs is given in~\citep{kim2018derivation}. An explanation for the non-uniqueness in
terms of the Gauge transformation appears in~\citep{abedi2019gauge}. An extension to the case with correlated noise appears in~\cite{kang2022optimal}. 

Given the non-uniqueness issue, a natural question is how to identify
a unique $\bar{X}$ process?  For this purpose, optimal transport
theory is described in the following~\Sec{sec:OT-FPF-path}.
For the linear Gaussian case, the theory is used to derive the
following optimal transport form of the linear FPF
(see~\citep{AmirACC2016,taghvaei2020optimal} for details):
\begin{equation}
	\begin{aligned}
		\ud \Xbar_t = &A \Xbar_t \ud t + \frac{1}{2}\Sigma_B\Sigmabar_t^{-1}\ (\Xbar_t - \mbar_t)\ud t \\
		&+\frac{1}{2}\Kbar_t(\ud Z_t - \frac{H \Xbar_t+H\mbar_t}{2}\ud t) + \Omega_t\Sigmabar_t^{-1}(\Xbar_t - \mbar_t)\ud t, 
	\end{aligned}\label{eq:opt-sde-Omega}
\end{equation}
where $\Omega_t=\Omega_t^{\text{OPT}}$ is a specific skew-symmetric matrix.    
The optimal transport FPF~\eqref{eq:opt-sde-Omega} is exact and has two differences compared to the
linear FPF~\eqref{eq:intro-FPF-Xbar-lin}:
\begin{enumerate}
	\item The stochastic term $\sigma_B\ud \Bbar_t$ is replaced with the deterministic term $\frac{1}{2}\Sigma_B \Sigmabar_t^{-1}(\Xbar_t-\mbar_t)\ud t$. Given a Gaussian prior, the two terms yield the same posterior.
	However, in a finite-$N$ implementation, the stochastic term serves to introduce an additional error of order $O(\frac{1}{\sqrt{N}})$~\citep[Prop. 4]{taghvaei2018error}.  
	\item The SDE~\eqref{eq:opt-sde-Omega} has an extra term involving the
	skew-symmetric matrix $\Omega_t$. The extra term does not effect the
	posterior, i.e., $\bar{X}$ is exact for {\em all}
        skew-symmetric choices of $\Omega_t$.  The specific optimal
        choice $\Omega_t=\Omega_t^{\text{OPT}}$ serves to pick the symmetric solution $G_t$ of the consistency
equation~\eqref{eq:G-gen}.
For the scalar
	($d=1$) case, the skew-symmetric term is zero.  Therefore, in the
	scalar case, the update formula in the linear FPF~\eqref{eq:intro-FPF-Xbar-lin} is optimal.  In
	the vector case, it is optimal iff $\Omega_t^{\text{OPT}}\equiv 0$.  
\end{enumerate} 

\subsection{FPF formula}
\label{sec:OT-FPF-path}

In this section, we provide a justification for the feedback control
formula in the FPF~\eqref{eq:intro-FPF-Xbar}. 
It is helpful to begin with the simpler deterministic
case.

\subsubsection{Deterministic path}
Let ${\mathcal P}_2(\Re^d)$ be the space of everywhere positive
probability densities on $\Re^d$ with finite second moment.  Given a smooth path
$\{p_t \in \mathcal P_2(\Re^d):~t\geq 0\}$ the problem is to
construct a stochastic process $\{\Xbar_t\}_{t\geq 0}$ such that the
probability density of $\Xbar_t$, denoted as $\bar{p}_t$, equals $p_t$ for all
$t\geq 0$. The exactness condition is expressed as
\begin{equation}\label{eq:exactness}
	\bar{p}_t = p_t,\quad \forall \;\; t\geq 0.
\end{equation}
As has already been noted, there are infinitely many stochastic
processes that satisfy the exactness condition.
A unique choice is made by
prescribing an additional optimality criterion based on the optimal
transport theory.  

To make these considerations concrete, assume that the given path
$\{p_t\}_{t\geq 0}$ evolves according to the PDE
\begin{equation*}
	\frac{\partial p_t}{\partial t} = \mathcal V(p_t),\quad t>0,
\end{equation*}
where $\mathcal V(\cdot)$ is an operator (e.g., the Laplacian) that acts
on probability densities.  (This necessarily restricts the
operator $\mathcal V$, e.g., $\int \mathcal V(\rho)(x) \ud x = 0$ for all
$\rho\in \mathcal P_2(\Re^d)$.) 
The following model is assumed for the process $\{\Xbar_t\}_{t\geq 0}$:
\begin{equation}\label{eq:Xbar-u}
	\frac{\ud}{\ud t} \Xbar_t  = u_t(\Xbar_t),\quad \Xbar_0 \sim p_0,
\end{equation}
where $u_t(\cdot)$ is a control law that needs to be designed. From the continuity equation, the exactness condition~\eqref{eq:exactness}
is satisfied if
\begin{equation}\label{eq:u-const}
	-\nabla \cdot(\bar{p}_t u_t) = \mathcal V(\bar{p}_t),\quad \forall\;\; t > 0.
\end{equation}

The non-uniqueness issue is now readily seen: The first-order
PDE~\eqref{eq:u-const} admits infinitely many solutions. 
A unique solution $u_t(\cdot)$ is picked by minimizing the
transportation cost from $\Xbar_t$ to $\Xbar_{t+\Delta t}$ in the
limit as $\Delta t \to 0$.  The $L^2$-Wasserstein cost is particularly
convenient because   
\begin{equation*}
	\lim_{\Delta t \to 0}\frac{1}{\Delta t^2}\mathbb{E}[|X_{t+\Delta t} - X_t|^2]=\int_{\Re^d} |u_t(x)|^2\bar{p}_t(x) \ud x.
\end{equation*}
Therefore, for each fixed $t$, the control law $u_t(\cdot)$ is
obtained 
by solving the constrained optimization problem 
\begin{equation*}\label{eq:opt-u}
	\min_{u_t(\cdot)} \int_{\Re^d} |u_t(x)|^2\bar{p}_t(x) \ud x,\quad \text{s.t}\quad    -\nabla \cdot(\bar{p}_t u_t) = \mathcal V(\bar{p}_t).
\end{equation*}
By a standard calculus of variation argument, the optimal solution is obtained as $u^*_t=\nabla
\phi_t$ where $\phi_t$ solves the Poisson equation $-\nabla
\cdot(\bar{p}_t \nabla \phi_t) = \mathcal V(\bar{p}_t)$. 
The resulting stochastic process $\Xbar$ is defined by
\begin{align*}
	&\frac{\ud \Xbar_t}{\ud t} = \nabla \phi_t(\Xbar_t),\quad \Xbar_0\sim p_0,\\
	&\phi_t \text{ solves the PDE } -\nabla \cdot(\bar{p}_t \nabla \phi_t) = \mathcal V(\bar{p}_t).
\end{align*}
The process is exact by construction.  

\begin{example}
Suppose the given path is a solution of the heat equation
$\frac{\partial p_t}{\partial t} = \Delta p_t$ ($\mathcal V(\cdot)$ is the
Laplacian).   The solution of the Poisson equation is easily
obtained as $\phi_t = \log(\bar{p}_t)$.  The optimal transport process
then evolves according to
\begin{subequations}
\begin{equation}\label{eq:Xbar_heat_det}
	\frac{\ud}{\ud t} \Xbar_t = -\nabla \log(\bar{p}_t(\Xbar_t)),\quad \Xbar_0\sim p_0.
\end{equation}
This process should be compared to the well known example
\begin{equation}\label{eq:Xbar-heat-s}
	\ud X_t = \ud B_t,\quad X_0\sim p_0,
\end{equation}
\end{subequations}
where $\{B_t\}_{t\geq 0}$ is a W.P..  The density for $X_t$ also
solves the heat equation.  In the language of optimal
transportation theory, the coupling defining~\eqref{eq:Xbar_heat_det}
is deterministic while it is stochastic in~\eqref{eq:Xbar-heat-s}.
\end{example}

\subsubsection{Stochastic path}
In the filtering problem, the path of the posterior probability
density is stochastic (because it depends upon the random
observations $\{Z_t\}_{t\geq 0}$). Therefore, the preceding discussion
is not directly 
applicable.  Suppose the stochastic path
$\{p_t\}_{t\geq 0}$ is governed by a
stochastic PDE
\begin{equation*}
	\ud p_t = \mathcal H(p_t) \ud I_t,
\end{equation*}
where $\mathcal H(\cdot)$ is an operator that acts on probability
densities 
and $\{I_t:t\geq 0\}$ is a W.P.. 

Consider the following SDE model:
\begin{equation*}
	\ud \Xbar_t = u_t(\Xbar_t)\ud t + \K_t(\Xbar_t) \ud I_t,\quad \Xbar_0 \sim p_0
\end{equation*}
where, compared to the deterministic model~\eqref{eq:Xbar-u}, 
an additional stochastic term is now included. The problem is to
identify control laws $u_t(\cdot)$ and $\K_t(\cdot)$ such that
the conditional density of $\Xbar_t$ equals $p_t$. Upon writing the
evolution equation for the conditional density of
$\Xbar_t$~\cite[Prop. 1]{yang2016}, the exactness condition is
formally satisfied by all such $u_t(\cdot)$ and $\K_t(\cdot)$ that solve
\begin{subequations}
	\begin{align}
		&-\nabla \cdot(\bar{p}_t\K_t) = \mathcal H(\bar{p}_t),\\
		&-\nabla \cdot(\bar{p}_t u_t) + 
		\frac{1}{2}(\nabla \cdot(\bar{p}_t\K_t) \K_t + \bar{p}_t\K_t\nabla \K_t) = 0.\label{eq:const-u}
	\end{align}
\end{subequations}
These equations are the stochastic counterpart of~\eqref{eq:u-const},
and as with~\eqref{eq:u-const}, their solution is not
unique.

The unique solution is obtained by requiring that the
coupling from $\Xbar_t$ and $\Xbar_{t+\Delta t}$ is optimal in the
limit as $\Delta t \to 0$.  
In contrast to the deterministic setting, the leading term in the
transportation cost $\mathbb{E}[|\Xbar_{t+\Delta t}- \Xbar_t|^2]$ is 
$O(\Delta t)$ whereby
\begin{equation*}\label{eq:opt-problem-K-1}
	\lim_{\Delta t \to 0}\frac{1}{\Delta t}\mathbb{E}[|\Xbar_{t+\Delta t} - \Xbar_t|^2]=\int_{\Re^d} |\K_t(x)|^2\bar{p}_t(x) \ud x.
\end{equation*}
Therefore, for each fixed $t$, the control law $\K_t(\cdot)$ is
obtained 
by solving the constrained optimization problem 
\begin{equation*}\label{eq:opt-problem-K-2}
	\min_{\K_t(\cdot)} \int_{\Re^d} |\K_t(x)|^2\bar{p}_t(x) \ud x,\quad \text{s.t}\quad    -\nabla \cdot(\bar{p}_t \K_t) = \mathcal H_t(\bar{p}_t).
\end{equation*}
As before, the optimal solution is given by 
$\K^*_t=\nabla \phi_t$ where $\phi_t$ solves the second-order PDE
$-\nabla \cdot(\bar{p}_t \nabla \phi_t) = \mathcal H(\bar{p}_t)$. 

It remains to identify the control law $u_t(\cdot)$.  For this purpose,
the second-order term in the infinitesimal Wasserstein cost is used: 
\begin{align*}
	\lim_{\Delta t \to 0}&\frac{1}{\Delta t^2}\left(\mathbb{E}[|\Xbar_{t+\Delta t} - \Xbar_t|^2] - \Delta t \int_{\Re^d} |\K^*_t(x)|^2\bar{p}_t(x)\ud x \right) \\&= \int_{\Re^d} |u_t(x)|^2\bar{p}_t(x) \ud x.
\end{align*}
The righthand-side is minimized subject to the
constraint~\eqref{eq:const-u}.  Remarkably, the optimal solution is
obtained in closed form as 
\[
u^*_t = -\frac{1}{2\bar{p}_t}\mathcal{H}(\bar{p}_t)\nabla \phi_t +
\frac{1}{2}\nabla^2 \phi_t \nabla \phi_t + \xi_t,
\] 
where $\xi_t$ is the (unique such) divergence free vector field
(i.e., $\nabla \cdot(p_t\xi_t)=0$) such that $u_t^*$ is of a gradient
form.  That~\eqref{eq:const-u} can be solved in an explicit manner was
a major surprise at the time of its discovery~\citep{yang2011mean,yang2013}. 
The resulting optimal transport process is
\begin{align}\label{eq:OT-process-stoch}
	&\ud \Xbar_t = \nabla \phi_t(\Xbar_t) \circ (\ud I_t - \frac{1}{2\bar{p}_t}\mathcal{H}(\bar{p}_t)\ud t) + \xi_t(\Xbar_t)\ud t ,~ \Xbar_0 \sim p_0.
	\end{align}
It is also readily shown that the process $\{\bar{X}_t\}_{t\geq 0}$ is
in fact exact for any choice of divergence free vector field
$\{\xi_t\}_{t\geq 0}$.  The most convenient such choice is to
simply set $\xi_t\equiv 0$.  The resulting filter is exact and
furthermore also (infinitesimally) optimal to the first-order.

For the special case of the nonlinear filtering problem,
$\mathcal{H}(\rho) = (h-\bar{h})\rho$ where $\bar{h} = \int h(x) \rho(x) \ud
x$ and $\ud I_t = (\ud Z_t - \bar{h}_t\ud t)$ is the increment of the
innovation process.  For these choices, the optimal transport stochastic process
\eqref{eq:OT-process-stoch} becomes 
\begin{align*}
	&\ud \Xbar_t = \nabla \phi_t(\Xbar_t) \circ (\ud Z_t -
          \frac{1}{2}(h(\Xbar_t) +\bar{h}_t)\ud t) + \xi_t(\Xbar_t)\ud
          t. 
\end{align*}
The feedback control law in the FPF algorithm~\eqref{eq:intro-FPF-Xbar}
represents the particular sub-optimal choice $\xi_t\equiv 0$.  The
choice is optimal for $d=1$.  

\revise{
\begin{remark}
The sub-optimality of FPF is not a problem because the filter is
exact.  A case for FPF may be made on computational grounds.  Because
it requires a solution of a single Poisson equation, the FPF
control law is the simplest possible control law leading to an exact
filter.  A natural question then is whether there is any advantage to
be had by using the optimal transport control law?  As of yet, the answer to
this question is not clear.  The same question arises in the optimal
transport map estimation  problem~\citep{makkuva2020optimal}: why aim
for the optimal transport map as opposed to say  Knothe--Rosenblatt
rearrangement~\citep[Ch. 1]{villani2009optimal}?  As an additional
point, there is also a freedom
in replacing the quadratic cost function in the optimal transport
problem. An argument for the optimal transport map with quadratic cost
function can be made on the account of its special geometrical
structure: the optimal map is the gradient of a convex function. This may lead to nice
computational and stability properties when the map is approximated
with particles/samples. 
\end{remark}}

\subsection{Optimal transport formula for the static
  example} \label{sec:OT-FPF-simpler-example}

We now revisit the static example introduced
in~\Sec{sec:FPF-simpler-example} with the aim of deriving an explicit
form of the control $U$ and relating it to the FPF.  As explained
in~\Sec{sec:FPF-simpler-example}, the problem is to find  a control
$U$ such that  $\mathbb E[f(X)|Y]= \mathbb E[f(\bar X_1)|Y]$ for all
functions $f\in C_b(\mathbb R^d) $, where $\bar X_1=\bar X_0+U$ and
$\bar X_0$ is an independent copy of  $X$.  This condition is equivalently expressed as $(\bar X_1,Y) \sim \mathsf P_{XY}$, and the problem of finding $U$ is formulated as the following optimal transportation problem:
\begin{equation}\label{eq:OT-formulation}
	\begin{aligned}
		&\min_{U \in \sigma(\bar X_0,Y)}~\mathbb E[|U|^2],\\\text{s.t} \quad &\bar X_1= \bar X_0 + U,\quad ( \bar X_1, Y) \sim \mathsf P_{XY},
	\end{aligned}
\end{equation}
where  the notation $U\in \sigma(\bar X_0,Y)$ means that $U$ is
allowed to be measurable with respect to  $\bar X_0$ and $Y$. This is
an optimal transportation problem between $(\bar X_0,Y) \sim \mathsf
P_X \otimes\mathsf  P_Y$ and $(X,Y) \sim \mathsf P_{XY}$ where the
transportation is constrained to be of the form $(\bar X_0,Y) \to
(\bar X_0 + U,Y)$, i.e., the second argument $Y$ remains fixed.  Its
solution is obtained as an extension of the celebrated
Brenier's result~\citep{brenier1991polar} as follows:

\begin{theorem}[Thm. 1 in
  \cite{taghvaei2022optimal}] \label{thm:OT-FPF} Consider the optimal
  transportation problem~\eqref{eq:OT-formulation}.  Suppose $\mathsf
  P_X$ admits a density with respect to the Lebesgue measure.  Then the
  optimal control is \[U = \nabla  \bar \Phi(\bar X_0;Y) - \bar X_0,\]
  where $\bar \Phi$ is the minimizer of the dual Kantorovich problem  
	\begin{align}\label{eq:dual-FPF}
		&\min_{\Phi \in \text{CVX}_x }~\mathbb E[\Phi(\bar X_0;Y) + \Phi^\star(X;Y)],
	\end{align} 
	where $\Phi \in \text{CVX}_x $ means $x \mapsto \Phi(x;y)$ is convex in $x$  for all $y$ and $\Phi^\star(x;y) := \sup_{z} z^\transpose x - \Phi(z;y)$ is the convex conjugate of $\Phi$ with respect to $x$. 
\end{theorem}
\begin{remark}[Relationship to the update formula for FPF]
In the continuous-time limit, the dual Kantorovich problem~\eqref{eq:dual-FPF} is related to the variational form~\eqref{eq:Poisson-opt} of the  Poisson equation~\eqref{eq:intro-Poisson}. 
In particular, with $\Delta Z_t=h(X_t) \Delta t + \Delta W_{t}$, the
solution to the problem~\eqref{eq:dual-FPF} is as follows~\cite[Prop. 2]{taghvaei2022optimal}:
		\begin{align*}
			\bar \Phi(\bar X_t;\Delta Z_t) = \frac{1}{2} |\bar
                  X_t|^2 + \phi (\bar X_t) \Delta Z_t + \psi(\bar X_t)\Delta t + O(\Delta t^2)
		\end{align*}
	where $\phi$ is the solution to the Poisson equation~\eqref{eq:intro-Poisson} with
        $\rho$ taken as the density of $\mathsf P_X$, and $\psi$ is the
        unique such function such that $\nabla \psi =  - \frac{h+\bar
          h}{2}\nabla \phi + \frac{1}{4}\nabla |\nabla \phi|^2  + \xi$
        where $\xi$ is divergence
        free. Therefore, the
        optimal transformation $\bar{X}_t\mapsto \bar{X}_{t+\Delta t}$ is
        given by,
	\begin{align*}
		\bar X_{t+\Delta t} &= \nabla_x \bar \Phi(\bar X_t;\Delta Z_t) \\&=
                                                                      \bar
                                                                      X_t
                                                                      +
                                                                      \nabla
                                                                      \phi
                                                                      (\bar
                                                                      X_t)
                                                                      (\Delta
                                                                      Z_t
                                                                      -\frac{h(\bar{X}_t)+\bar
                                                                      h_t}{2}\Delta
                                                                      t)           \\&\qquad\;\; 
                                                                      +
                                                                      \frac{1}{4}\nabla
                                                                      |\nabla
                                                                      \phi(\bar
                                                                      X_t)|^2
                                                                      \Delta
                                                                      t
+ \xi(\bar X_t)\Delta t + O(\Delta t^2)
	\end{align*}  
which in the limit as $\Delta t \to 0$ is the SDE for the optimal transport
FPF~\eqref{eq:OT-process-stoch}. 
\end{remark}

\begin{remark}[Stochastic optimization and DNNs]
The variational problem~\eqref{eq:dual-FPF} is a
stochastic optimization problem which allows for application of machine learning
tools to approximate its solution. In particular, deep neural networks
(DNNs) can be used to parameterize 
the function  $\Phi$ and stochastic optimization algorithms employed to learn the parameters. Preliminary results in this direction are
presented in~\citep{taghvaei2022optimal} with a comprehensive
development the subject of ongoing work. 
\end{remark}

\medskip

\begin{center}
	PART II
\end{center}
\section{CIPS for optimal control}
\label{sec:LQR}

In order to elucidate the ideas as clearly as possible, our focus in
this paper is entirely on the linear
quadratic (LQ) problem.  Its extension to the
nonlinear optimal control
problem~\eqref{eq:nonlinear_opt_control_problem} can be found in \citep{joshi2022controlled}.

\subsection{Problem statement and background}

The finite-horizon linear quadratic (LQ) optimal control problem is a
special case of~\eqref{eq:nonlinear_opt_control_problem} as follows:

\begin{subequations} 
	\label{eq:LQ}	
	\begin{align}
		\min_{u} \quad J(u) &= \int_0^T \half \left(|C x_t|^2 + u_t^\transpose \RU u_t\right) \ud t + \ x_T^\transpose
		P_T x_T  \label{eq:LQ:model-objective} \\
		\text{subject to:} \quad \dot{x}_t &= A x_t + B u_t, \quad x_0=x \label{eq:LQ:model-dynamics}
	\end{align} 
\end{subequations}
It is assumed that $(A,B)$ is controllable, $(A,C)$ is observable,
and matrices $P_T,\RU \succ 0$.  The $[T=\infty]$ limit is referred to as the linear quadratic regulator (LQR)
problem.   

It is well known
that the optimal control $u_t = \fee_t(x_t)$ where the optimal policy is linear
\[
\fee_t(x)=K_t x \quad \text{where}\quad K_t = - \RU^{-1}B^\transpose
P_t,\quad 0\leq t\leq  T
\]
is the optimal gain matrix and $\{P_t:0\leq t\leq T\}$ is a solution of the backward (in time) {differential Ricatti equation (DRE)}
\begin{equation}\label{eq:Ricatti}
	-\frac{\ud}{\ud t} P_t =A^\transpose P_t + P_t A + C^\transpose C  - P_tB  \RU^{-1}B^\transpose P_t,
	\quad P_T \;\text{(given)}
\end{equation}
The {algebraic  Ricatti equation (ARE)} is obtained by setting the left-hand side to $0$.  As $T\to\infty$, 
for each fixed time $t$, $P_t \to P^{\infty}$, exponentially fast
\cite[Thm.~3.7]{kwakernaak1972linear}, where $P^{\infty} \succ 0$ is the unique
such positive-definite solution of the ARE, and therefore the optimal gain converges,
$K_t \to K^{\infty}:=- \RU^{-1}B^\transpose P^{\infty}$.   Approximation of the gain $K^{\infty}$ is
a goal in recent  work on model-based RL for the LQR problem~\citep{fazel_global_2018,mihailo-2021-tac}.

\subsection{Objectives and assumptions}

For the reasons noted in~\Sec{sec:intro}, we are interested in a
simulation-based solution that does not rely on an explicit solution
of the DRE~\eqref{eq:Ricatti}.  To clarify what is meant by a
simulation-based solution in the context of model-based RL, we make a formal assumption as follows:
\begin{assumption}
	\label{ass:Ass1}
	\begin{enumerate}
		\item Functions $f(x,\alpha)=A x+ B \alpha$ and $c(x)=C x$ are
		available in the form of an oracle (which allows function
		evaluation at any state action pair $(x,\alpha)\in\Re^d\times \Re^m$).
		\item Matrices $R$ and $P_T$ are available.  Both of these matrices
		are strictly positive-definite.
		\item Simulator is available to simulate~\eqref{eq:LQ:model-dynamics}.  
		\item Simulator provides for an ability to add additional inputs
		outside the control channel (e.g.,~see~\eqref{eq:CIPS_intro}).
	\end{enumerate}
\end{assumption}
This assumption is motivated from the data assimilation literature
where it is entirely standard and widely used in applications, such
as weather prediction, involving EnKF. Part 1 of the assumption means
that the matrices $A,B,C$ are not available explicitly.  Rather, for
any given $(x,\alpha)\in\Re^d\times \Re^m$, the vectors $f(x,\alpha)$
and $c(x)$ can be evaluated.   Function evaluation forms for the
dynamics and the cost function is also a standard assumption for any
model-based RL algorithm.  
Part 2 of the assumption is
not too restrictive for the following two reasons: 
\begin{enumerate}
	\item In physical systems,
	one is typically able to assess relative costs for different control
	inputs (actuators).  This knowledge can be used to select $R$.
	\item For the LQR problem, under mild technical
	conditions, the optimal policy is stationary and does not depend upon
	the choice of $P_T$.
\end{enumerate}
If these matrices are not available, one possibility is to take $R$ and $P_T$ to
be identity matrices of appropriate dimensions.  
The main restriction comes from part 3 of the assumption.  However, as
the widespread use of EnKF amply demonstrates, it is not un-realistic
to assume it for a simulation-based solution.  Of course, it will
not be possible with a physical experiment.

\subsection{Dual EnKF}\label{sec:dual-EnKF}

The dual EnKF algorithm is obtained from making use of duality between optimal control
and filtering.  For this purpose, we need to first
dualize the DRE~\eqref{eq:Ricatti}.  
Under the assumptions of this paper, $P_t\succ 0$ for $0\leq t\le T$ whenever $P_T\succ
0$~\cite[Sec.~24]{brockett2015finite}.
Set $S_t = P_t^{-1}$.  It is readily verified that $\{S_t:0\leq t\leq T\}$ also solves a
DRE (which represents the dual of~\eqref{eq:Ricatti})
\begin{equation}\label{eq:backward-DRE}
	\frac{\ud}{\ud t} S_t = AS_t + S_t A^\transpose - B  \RU^{-1}B^\transpose +
	S_tC^\transpose C S_t,\quad S_T=P_T^{-1}
\end{equation}     

The strategy is to approximate $\{S_t:0\leq t\leq T\}$ using simulations. As before, the construction proceeds
in two steps: (i) definition of an exact mean-field process; and (ii) its finite-$N$
approximation.

\medskip

\noindent \textbf{Step 1. Mean-field process:} 
Define a stochastic process $\Ybarbar=\{\Ybarbar_t\in \Re^d:
0\leq t\leq T\}$ as a solution of the
following backward (in time) SDE:
\begin{align}
	\ud \Ybarbar_t & = A \Ybarbar_t \ud t   +  B\ud  \backward{\etabar}_t
	+\half \Sbar_tC^\transpose (C\Ybarbar_t +
	C \bar{n}_t) \ud t,  \;0\leq t<T\nonumber \\ \Ybarbar_T & \sim
	\mathcal
	N(0,S_T) \label{eq:Ybar} 
\end{align}
where ${\etabar}=\{{\etabar}_t \in \Re^m:0\leq t\leq T\}$ is a W.P. 
with covariance matrix $\RU^{-1}$, and 
\begin{align}
	\bar{n}_t:=\mathbb{E} [\Ybarbar_t], \quad \Sbar_t
	:=\mathbb{E} [(\Ybarbar_t-\bar{n}_t)(\Ybarbar_t-\bar{n}_t)^\transpose],  \;0\leq t<T \label{eq:Sbar}
\end{align}
The meaning of the backward
arrow on $\ud  \backward{\etabar}$ in~\eqref{eq:Ybar} is that the SDE
is simulated backward in time starting from the terminal condition
specified at time $t=T$.  The reader is referred
to~\citep[Sec. 4.2]{nualart1988stochastic} for the definition of the
backward It\^{o}-integral. 
The mean-field process is useful because of the following proposition.
\medskip

\begin{proposition}[Prop. 1 in \cite{joshi2022controlled}]\label{prop:Y-exactness}
	The solution to 
	the SDE~\eqref{eq:Ybar}
	is a Gaussian stochastic process, in which the mean and covariance of $\Ybarbar_t$ are given by
	\[
	\bar{n}_t = 0,\quad \bar{S}_t=S_t, \quad \;
	0\leq t\leq T
	\]
	Consequently, $\Ybar_t  := \Sbar_t^{-1} (\Ybarbar_t-\bar{n}_t)$ is also
	a Gaussian random variable with 
	\[
	\mathbb{E} [\Ybar_t] = 0,\quad \mathbb{E} [\Ybar_t \Ybar_t^\transpose]= P_t, \quad\;
	0\leq t\leq T
	\]
\end{proposition}

The significance of~\Prop{prop:Y-exactness} is that the optimal control
policy $\fee_t(\cdot)$ can now be obtained in terms of the statistics of the random variable
$\Ybar_t$.  Specifically, we have the following two cases: 
\begin{enumerate}
	\item If the matrix $B$ is explicitly known then the optimal gain matrix 
	\[
	K_t =  - \RU^{-1}B^\transpose \mathbb{E} [\Ybar_t
	\Ybar_t^\transpose]
	\]
	\item If $B$ is unknown, define the Hamiltonian (the
	continuous-time counterpart of the Q-function~\citep{mehta2009q}):
	\begin{align*} & H(x,\alpha,t) \\
		&\;\;:= \underbrace{\half
			|Cx|^2 + \half \alpha^\transpose R \alpha}_{\text{cost function}} +
		x^\transpose \mathbb{E}  [\Ybar_t \Ybar_t^\transpose]\underbrace{(Ax + B\alpha)}_{\text{model}~\eqref{eq:LQ:model-dynamics}} 
	\end{align*}
	from which the
	optimal control law is obtained as
	\[
	\fee_t(x) = \argmin_{\alpha\in\Re^m} H(x,\alpha,t)
	\] 
	by recalling the minimum principle, which states that the optimal control is the unique minimizer of the
	Hamiltonian. It is noted that the Hamiltonian $H(x,\alpha,t)$ is in the form of an oracle because $(Ax
	+ B\alpha)$ is the right-hand side of the simulation
	model~\eqref{eq:LQ:model-dynamics}.
\end{enumerate}

\medskip

\noindent \textbf{Step 2. Finite-$N$ approximation:} 
The particles $\{Y^i_t \in\Re^d: 0 \leq t\leq T, i =1,\ldots,N\}$ evolve
according to the backward SDE:
\begin{align}
	\ud {Y}^i_t & = \underbrace{A {Y}^i_t \ud t + B\ud \backward{\eta}^i_t}_{\text{i-th copy of model}~\eqref{eq:LQ:model-dynamics}} +
	\underbrace{S^{(N)}_tC^\transpose
		\left(\frac{C{Y}^i_t+Cn^{(N)}_t}{2}
		\right)}_{\text{coupling}}\ud
	t, \label{eq:EnKF-Yi} \\
	{Y}^i_T & \stackrel{\text{i.i.d}}{\sim} \mathcal
	N(0,P_T^{-1}),\quad 1\leq i\leq N \nonumber
\end{align}
${\eta}^i:=\{\eta_t^i:0\leq t\leq T\}$ is an i.i.d copy of ${\etabar}$
and
\begin{align*}
n^{(N)}_t &= \frac{1}{N}\sum_{i=1}^N
{Y}^i_t \\
\SN_t & = \frac{1}{N-1}\sum_{i=1}^N
	({Y}^i_t-n^{(N)}_t)({Y}^i_t-n^{(N)}_t)^\transpose \label{eq:SbarN}
\end{align*}
The CIPS~\eqref{eq:EnKF-Yi} is referred to as the {\em
	dual EnKF}.  
\medskip

\noindent \textbf{Optimal control:} Set $
X^i_t  = (\SN_t)^{-1} ({Y}^i_t  - n^{(N)}_t)$.  
There are two cases as before: 
\begin{enumerate}
	\item If
	the matrix $B$ is explicitly known then
	\begin{equation}
			K_t^{(N)}  = - \frac{1}{N-1} \sum_{i=1}^N \RU^{-1}
		(B^\transpose  X^i_{t} )(X^i_{t})^\transpose 
		\label{eq:KtN}
	\end{equation}
	\item  If $B$ is unknown, define the Hamiltonian 
	\begin{align*}
		H^{(N)}(x,\alpha,t) :=&
		\underbrace{\half
			|Cx|^2 + \half \alpha^\transpose R \alpha}_{\text{cost function}} \\&+
		\frac{1}{N-1} \sum_{i=1}^N  (x^\transpose  X^i_{t} )(X^i_{t})^\transpose \underbrace{(Ax + B\alpha)}_{\text{model}~\eqref{eq:LQ:model-dynamics}} 
	\end{align*}
	from which the
	optimal control policy is approximated as
	\[
	\fee_t^{(N)}(x) = \argmin_{a\in\Re^m} H^{(N)}(x,a,t)
	\]
	There are several zeroth-order approaches to solve the minimization problem, e.g., by
	constructing 2-point estimators for the
	gradient.  Since the
	objective function is quadratic and the matrix $R$ is known, $m$ queries
	of $H^{(N)}(x,\cdot,t)$ are sufficient to compute $\fee_t^{
		(N)}(x)$. 
	
\end{enumerate}

\begin{figure*}[t]
	\centering
		\subfigure[$d=2$]{
			\centering
			\includegraphics[width=1\columnwidth]{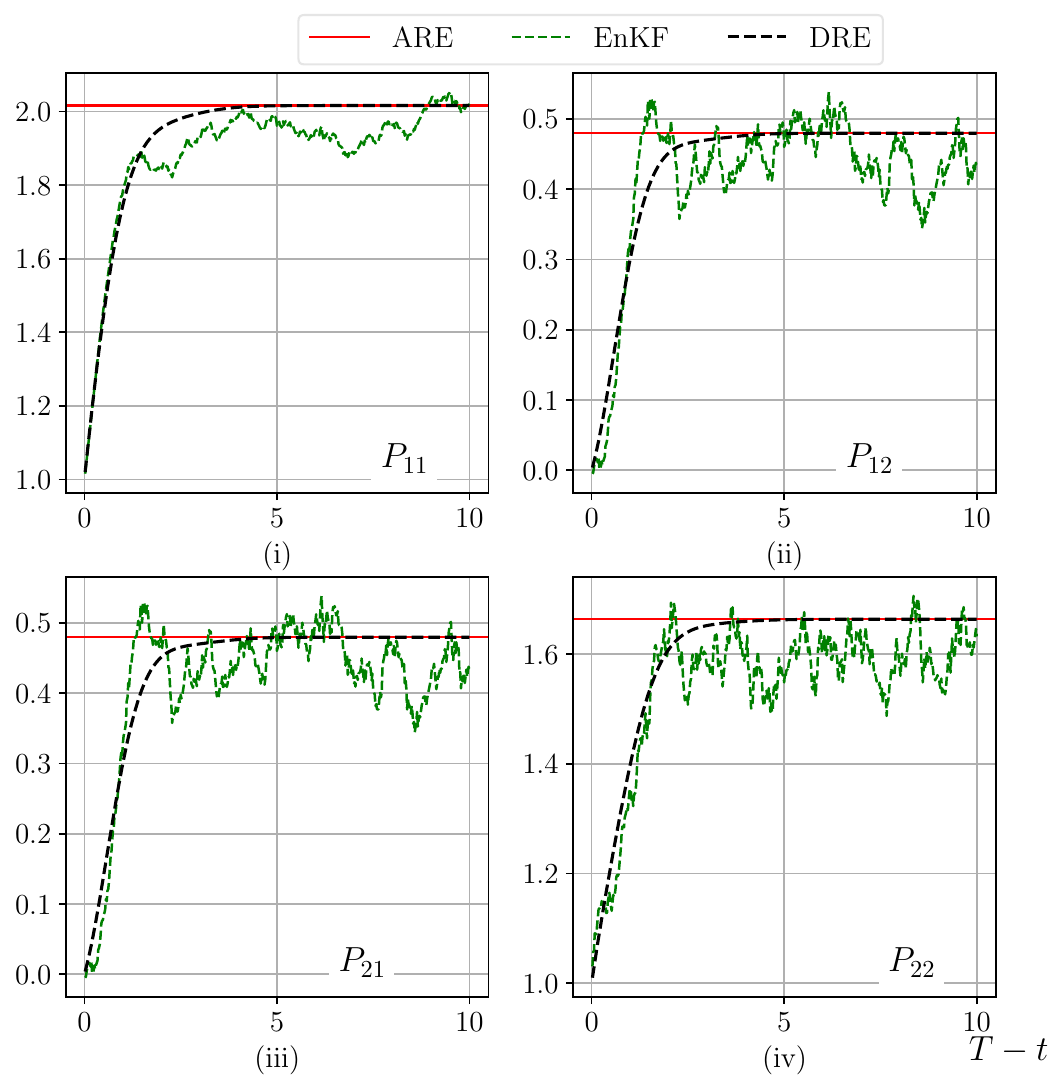}  
			\label{fig:enkf-learns-4}
		}
	\hfill
		\subfigure[$d=10$]{
			\centering
			\includegraphics[width=1\columnwidth]{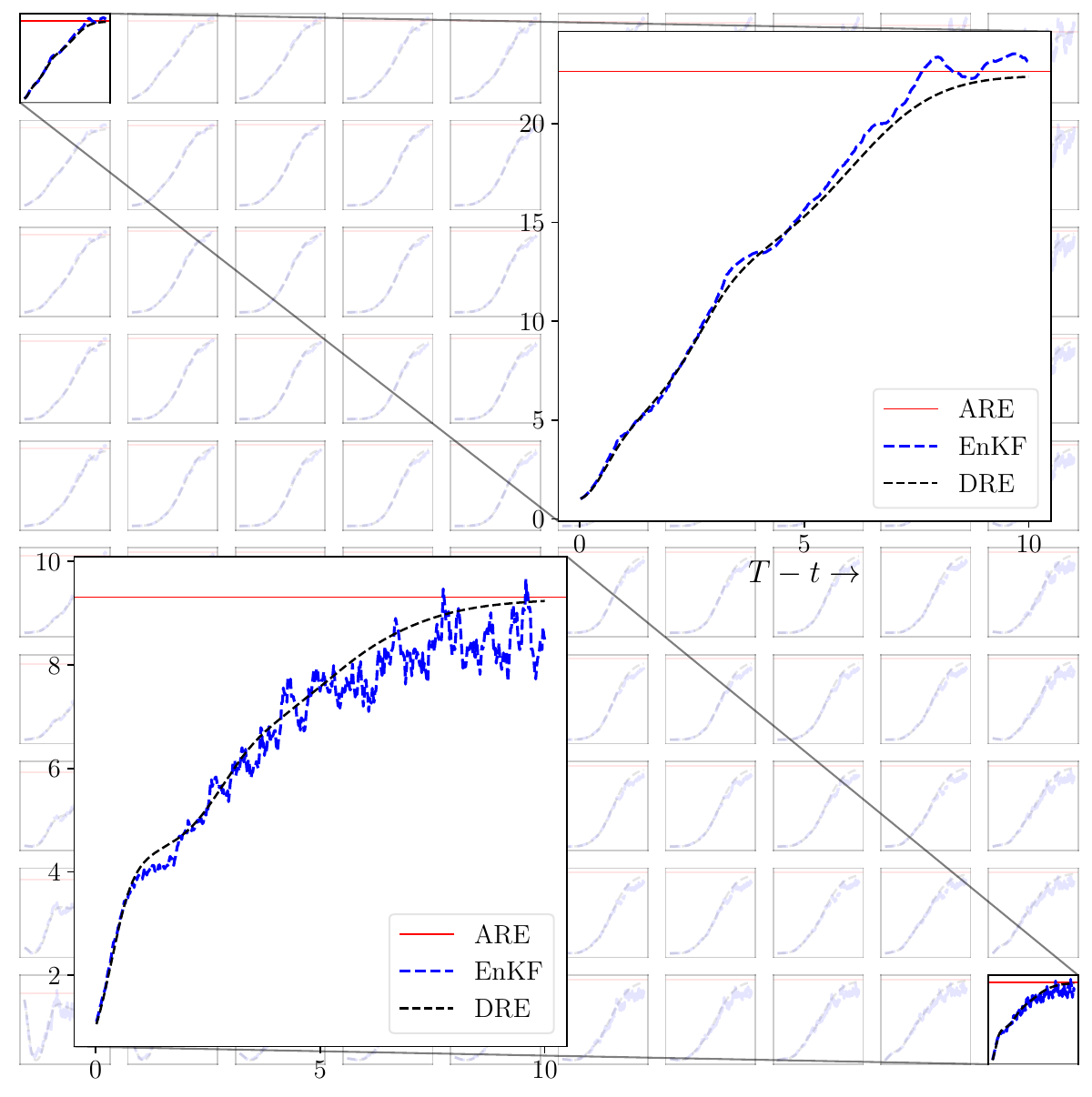}  
			\label{fig:enkf-learns-100}
		}
	\caption{Comparison of the numerical solution obtained
		from the EnKF, the DRE, and the ARE. Note the $x$-axis
		for these plots is $T-t$ for $0\leq t\leq T$. $d$ is
                the state-dimension. }
	\label{fig:convergence}
	\vspace{-0.1in}
\end{figure*}

The overall dual EnKF algorithm can be found in~\citep[Algorithm 1 and 2]{joshi2022controlled}.

\subsection{Relating dual EnKF to model-based RL}
\label{sec:dual-EnKF-explanation}
The following remarks are included to help provide an intuitive
explanation of the various aspects of the dual EnKF and relate
these to the model-based RL:

\begin{enumerate}
\item{\textbf{Representation}.}
In designing any RL algorithm, the
first issue is the representation of the unknown
value function ($P_t$ in the linear case).  Our novel idea is to represent $P_t$ is in terms of statistics
(variance) of the particles.  Such a representation is 
distinct from representing the value function, or its proxies, such as
the Q function, within a parameterized class of functions. 

\item{\textbf{Value iteration}.}
The algorithm is entirely simulation based: $N$ copies of the
model~\eqref{eq:LQ:model-dynamics} are simulated in parallel where the
terms on the right hand-side of~\eqref{eq:EnKF-Yi} have the following
intuitive interpretations:

\begin{enumerate}
	\item {\em Dynamics:} The first term ``$A {Y}^i_t \ud t$'' on the right-hand side of~\eqref{eq:EnKF-Yi} is
	simply a copy of uncontrolled dynamics in the
	model~\eqref{eq:LQ:model-dynamics}.
	
	\item {\em Control:} The second term ``$B\ud
          \backward{\eta}^i_t$'' is 
	the control input for the $i$-th particle.  It is specified as a W.P. with covariance $\RU^{-1}$.  One may interpret this as an approach to exploration whereby cheaper control
	directions are explored more.   

\item {\em Coupling:} The third term, referred to as the coupling,
  effectively implements the value iteration step.  Coupling has a ``gain times error'' structure where $S^{(N)}_tC^\transpose$ is the gain and $\frac{1}{2} (C
Y_t^i + Cn_t^{(N)})$ is the counterpart of the error in the
linear FPF~\eqref{eq:intro-FPF-Xbar-lin}. 
\end{enumerate}

\item{\textbf{Arrow of time}.}
The particles are simulated
backward---from terminal time $t=T$ to initial time $t=0$.  This is
different from most model-based RL but 
consistent with the dynamic programming (DP) equation which
also proceeds backward in time.  

\end{enumerate}

\vspace*{-0.02in}
\subsection{Convergence and error analysis}\label{sec:dual-EnKF-conv-anal}
\vspace*{-0.02in}

In \citep[Prop. 3]{joshi2022controlled}, under
certain additional assumptions on system matrices, the following
error bound is derived:
\begin{equation}\label{eq:error_formula}
	\mathbb{E}[\|S^{(N)}_{t}-\bar{S}_t\|_F] \leq
	\frac{C_1}{\sqrt{N}} + C_2e^{-2\lambda (T-t)} \mathbb{E}[\|\SN_T-\Sbar_T\|_F],
\end{equation}
where $C_1,C_2,\lambda$ are positive constants and $||\cdot||_F$ denotes Frobenius norm for matrices.
The significance of the bound~\eqref{eq:error_formula} is as follows:
The constant $\lambda$ is same as the rate that governs the
convergence of the solution of the DRE~\eqref{eq:Ricatti} to the
stationary solution (of the
infinite-horizon LQR problem).  {\em This means that the dual EnKF learns
the optimal LQR gain exponentially fast with a rate
that is as good as one would obtain from directly solving the DRE}.

Convergence is numerically illustrated for a d-dimensional system
expressed in its controllable canonical form 
\begin{equation*}
	A = \begin{bmatrix}
		0&1&0&0&\ldots&0\\
		0&0&1&0&\ldots&0\\
		\vdots & & & & & \vdots\\
		a_1 & a_2 & a_3 & a_4 &\ldots & a_d
	\end{bmatrix},\quad 
	B = \begin{bmatrix}
		0\\
		0\\
		\vdots \\
		1 
	\end{bmatrix}
\end{equation*}
where the entries $(a_1,\ldots,a_d)\in \mathbb{R}^d$ are i.i.d.
samples from ${\cal N}(0 ,1)$.  The matrices $C,R,P_T$ are identity
matrices of appropriate dimension.  For numerics, we fix $T=10$, chose the
time-discretization step as $0.02$, and use $N=1000$ particles to
simulate the dual EnKF.

\begin{figure*}[t]
	\centering
		\subfigure[$d=2$]{
			\centering
			\includegraphics[width=0.9\columnwidth]{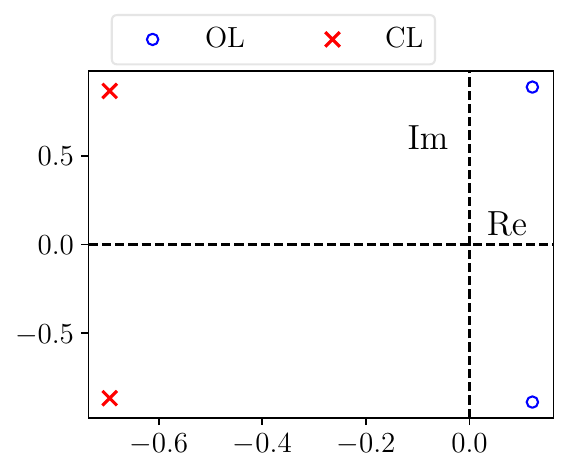}  
			\label{fig:poles-2}
		}
		\hfill
			\subfigure[$d=10$]{
				\centering
				\includegraphics[width=0.9\columnwidth]{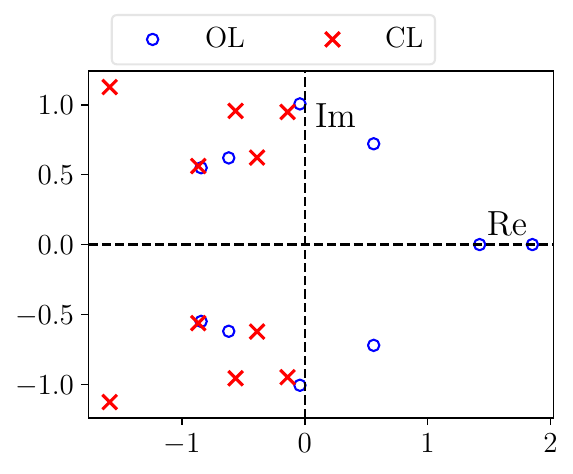}  
				\label{fig:poles-10}
			}
		\caption{Open and closed-loop poles for the two plots
                  (parts (a) and (b))
                  depicted in \Fig{fig:convergence}.}
		\label{fig:poles}
		\vspace{-0.1in}
	\end{figure*}

\Fig{fig:enkf-learns-4} depicts the convergence of the four
entries of the matrix $P_t^{(N)}$  for the case where $d=2$.  
\Fig{fig:enkf-learns-100} depicts the analogous results for
$d=10$. \Fig{fig:poles-2} and \Fig{fig:poles-10} depict the
open-loop poles (eigenvalues of the matrix $A$) and the closed-loop
poles (eigenvalues of the matrix $(A + BK_0^{(N)})$), for $d=2$ and $d=10$,
respectively. Note that the closed-loop poles are stable, whereas some
open-loop poles have positive real parts.

	\subsection{Comparison to literature}
	\label{sec:dual-EnKF-comparison}
	
	We present a comparison of the dual EnKF with policy
	gradient algorithms in~\cite{mihailo-2021-tac} (denoted as
	[M21]) and~\cite{fazel_global_2018} (denoted as [F18]).  In
        these prior works, by restricting the control policies to the
        linear form $u_t = Kx_t$, the LQR problem 
        reduces to the finite-dimensional static optimization problem: 
	\begin{align}
		K^\star =\argmin_K
		J(K) = \mathbb{E} \left(\int_{0}^{\infty} x_t^\transpose Qx_t + u_t^\transpose Ru_t \,
		dt \right)
		\label{eq:mihailo-LQ-cost} 
	\end{align}
	where the expectation is over the initial condition.    The
        authors apply a pure-actor method using ``zeroth order''
        methods to approximate gradient descent, much like the early
        REINFORCE algorithm for RL \citep{sutton-barto}.

A  qualitative comparison of the dual EnKF with these prior
algorithms is given in
Table~\ref{tab:EnKF-comparison}.  Choosing $t=0$ in~\eqref{eq:error_formula}, the error is
smaller than $\varepsilon$ if the number of particles
$N>O(\frac{1}{\varepsilon^2})$ and the simulation time
$T>O(\log(\frac{1}{\varepsilon}))$, while the iteration number is
one. This is compared with policy optimization approach
in~\cite{fazel_global_2018} where the number of particles and the
simulation time scales polynomially with $\varepsilon$, while the
number of iterations scale as $O(\log(\frac{1}{\varepsilon}))$. This
result is later refined in~\cite{mihailo-2021-tac} where the required
number of particles and the simulation time are shown to be $O(1)$ and
$O(\log(\frac{1}{\varepsilon}))$ respectively (although this result is
valid with probability that approaches zero as the number of
iterations grow~\cite[Thm. 3]{mihailo-2021-tac}.).

	\renewcommand{\arraystretch}{1.2}
	\begin{table*}[h]
		\centering 
		\begin{tabular}{|c|c|c|c|}
			\hline 
			Algorithm & particles/samples & simulation time & iterations \\\hline
			dual EnKF & $O(\frac{1}{\varepsilon^2})$  &  $O(\log(\frac{1}{\varepsilon}))$ &  $1$ \\ \hline
			\cite{fazel_global_2018} & $\text{poly}\left(\frac{1}{\varepsilon}\right)$ &  $\text{poly}\left(\frac{1}{\varepsilon}\right)$ &  $O(\log(\frac{1}{\varepsilon}))$ \\ \hline
			\cite{mihailo-2021-tac}  & $O(1)$  &  $O(\log(\frac{1}{\varepsilon}))$ &  $O(\log(\frac{1}{\varepsilon}))$   \\\hline
		\end{tabular}
		\caption{Computational complexity comparison of the algorithms to achieve $\varepsilon$ error in approximating the infinite-horizon LQR optimal gain. }
		\label{tab:EnKF-comparison}
	\end{table*}
	\renewcommand{\arraystretch}{1}

\begin{figure*}[h]
		\centering
		
		\includegraphics[width=2\columnwidth]{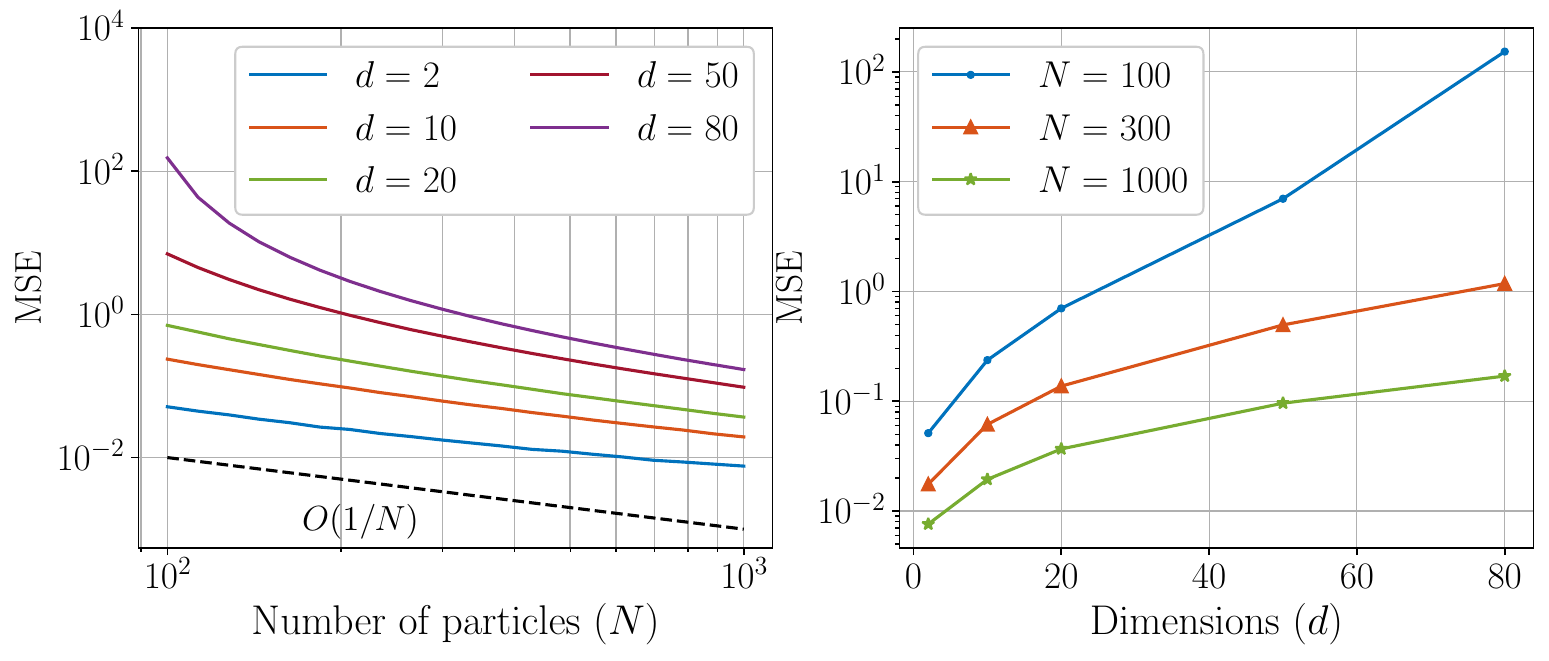}  
		
		\caption{Performance of the dual EnKF algorithm: MSE as a function of the
			number of particles $N$ and system dimension $d$.}
		\label{fig:mse}
	\end{figure*}

A  numerical comparison is 
	made on the benchmark spring mass damper example borrowed
        from~\citep[Sec.~VI]{mohammadi_global_2019}.  \Fig{fig:mse}
        depicts the relative mean-squared error, defined as
	\[
	\text{MSE} := \frac{1}{T} \mathbb{E}\left( \int_{0}^{T} \frac{\|
		P_t - P_t^{(N)} \|_F^2}{\| P_t \|_F^2} \: \ud t \right)
	\]  
	Two trends are depicted in the figure: the
	$O(\frac{1}{N})$ decay of the MSE as $N$ increases (for $d$
        fixed), which is a numerical illustration of the error bound \eqref{eq:error_formula},
	and a plot of the MSE as a function of dimension $d$ (for $N$ fixed).
	
\begin{figure*}[h]
		\centering
		
		\includegraphics[width=2\columnwidth]{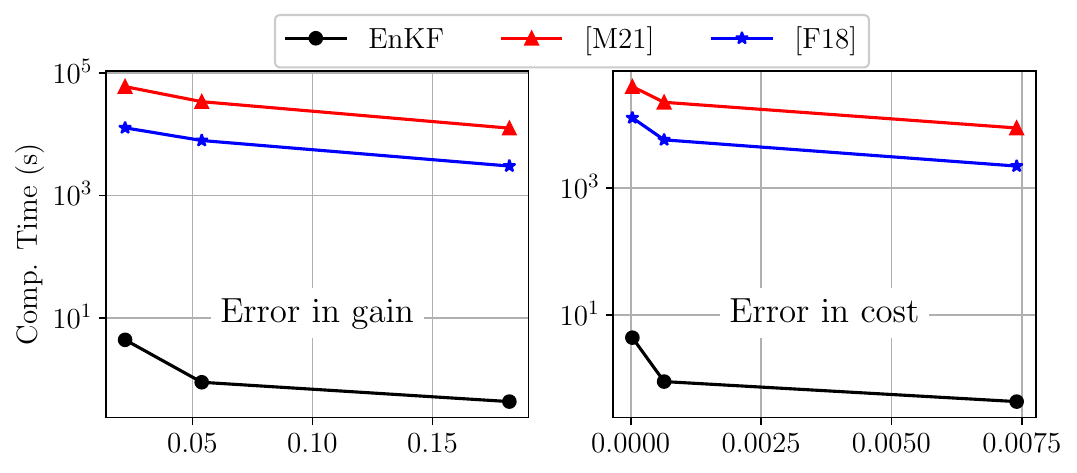}  
		
		\caption{Comparison with algorithms
			in~\cite{fazel_global_2018} (labeled [F18])
			and~\cite{mihailo-2021-tac} (labeled [M21]).  The
			comparisons depict the computation time (in Python) as a
			function of the relative error in approximating the LQR gain and cost.}\label{fig:comp-to-lit}
	\end{figure*}

	A side-by-side comparison with [F18] and [M21] is depicted in
	Fig.~\ref{fig:comp-to-lit}.  The comparison is for the following
	metrics (taken from \cite{mihailo-2021-tac}):
	\[
	\text{error}^{\text{gain}}= \frac{\| K^{\text{est}} - K^{\infty} \|_F}{\|K^{\infty}\|_F} \, ,
	\quad \text{error}^{\text{value}} =\frac{ c^{\text{est}} - c^{\infty} }{c^{(N)}_{\text{init}} - c^{\infty} }
	\]
	where the LQR optimal gain $K^{\infty}$ and the optimal value
	$c^{\infty}$ are computed from solving the ARE.  The value
	$c^{(N)}_{\text{init}}$ is approximated using the initial gain $K=0$
	(Note such a gain is not necessary for EnKF).  Because [F18]
	is for discrete-time system, an Euler approximation is used to obtain
	a discrete-time model.  
	
	In the numerical experiments, the dual EnKF is found to be
	significantly more
	computationally efficient---by two orders of magnitude or more. 
	The main reason for the order of
	magnitude improvement in computational time is as follows:  An EnKF
	requires only a single iteration over a fixed time-horizon
	In contrast, [F18] and [M21] require several steps of gradient
        descent, with each step requiring an evaluation of the LQR
        cost, and because these operations must be done serially,
        these computations are slower.

	In carrying out these comparisons, the same time-horizon $[0,T]$ and discretization
	time-step $\Delta t$  was used for all the algorithms.  It is certainly
	possible that some of these parameters can be optimized to improve the
	performance of the other algorithms. In particular, one may consider
	shorter or longer time-horizon $T$ or use parallelization to speed up the gradient calculation.  Codes are made
	available on Github for interested parties to independently verify
	these comparisons\footnote{\url{https://github.com/anantjoshi97/EnKF-RL}}.

	{
\subsection{Extension to the nonlinear problem~\eqref{eq:nonlinear_opt_control_problem}}
An extension of the dual EnKF algorithm for the nonlinear
optimal control problem~\eqref{eq:nonlinear_opt_control_problem}
appears in~\cite[Sec. 3]{joshi2022controlled}. In the general
nonlinear setting, the empirical distribution of the $N$ particles
approximates the minus log of the value function, leading to the optimal control
law~\eqref{e:policy}.  The algorithm involves the solution of a
Poisson equation, similar to the Poisson equation that appears in the
FPF algorithm. The dual EnKF algorithm for
the LQ problem arises as a special case when the Poisson equation admits an
analytical solution.  An interested reader can find additional details
in~\citep{joshi2022controlled} where some numerical results
for the problem of stabilizing an inverted pendulum on the cart are also
described.

}
	
\section{Discussion and conclusion}
\label{sec:conc}

In this survey, we described {CIPS} to approximate the solution of the optimal filtering and the
optimal control problems (in parts I and II, respectively).  As
explained in \Sec{sec:intro}, there are close parallels with {DA and RL}.  In this section,
we expand on some of these parallels with the goal of highlighting
some important points and directions for future work.

\paragraph{1. Data
assimilation, sampling, optimal transportation} CIPS may be viewed as
a sampling algorithm.  The FPF control law (coupling) is designed to
sample from the posterior.  Compared to the conventional particle
filters, coupling is beneficial because the issue of particle degeneracy is
avoided (as discussed in~\Sec{sec:FPF_comp}).   To design the coupling, optimal
transportation theory provides a useful framework (as described
in~\Sec{sec:OT-FPF}).  
Variations of the basic approach described here have been
used in construction of a class of filtering
algorithms~\citep{halder2017gradient,halder2018gradient,halderproximal,garbuno2020interacting,luo2019multivariate}.  
The optimal transport formulation has also been extended to the
Schr\"odinger bridge setting by considering a cost with respect to the
(prior) dynamics, or considering an entropic
regularization~\citep{chen2016relation,reich2019data}.  
In related works, the coupling viewpoint along with
geometric notions from optimal transportation theory, have enabled
application of optimization algorithms to design sampling
schemes~\citep{liu2016stein,richemond2017wasserstein,zhang2018policy,frogner2018approximate,chizat2018global,chen2018unified,liu2018accelerated,zhang2019mean,taghvaei2019accelerated}.

Part II of
this paper is
motivated by the enormous success of the CIPS (EnKF) in DA.

\paragraph{2. Reinforcement learning and optimal control} Compared to typical RL approaches, there are two
key innovations/differences: 
\begin{enumerate}
\item Representation of the unknown
value function in terms of the statistics (variance) of a suitably
designed process; and 
\item Design of interactions (coupling) between simulations
for the purposes of policy optimization.  
\end{enumerate}
We fully believe that the two key innovations may be useful for many
other types of models including MDPs and partially observed problems.
In the LQ setting of the problem, doing so is beneficial because of
the learning rate: Since the $[N=\infty]$ limit is
exact tor the LQ problem, the dual EnKF algorithm yields a learning
rate that closely approximates the exponential rate of convergence of
the solution of the DRE.  This is rigorously established with the aid
of error bound~\eqref{eq:error_formula}.  In numerical examples, this property is shown to lead to an 
order of magnitude better performance than the state-of-the-art
algorithms.  

Apart from RL, model predictive control (MPC) is another area where a model
in the form of a simulator is assumed to design optimal control for
problems such as~\eqref{eq:nonlinear_opt_control_problem}~\citep{rawlings2017model}.  Using
duality, MPC methods have been adapted to design the moving horizon
estimator (MHE).  A big selling point of MPC is its ability to handle
constraints which has not been a major theme in the DA literature.  
Another notable distinction is that while MPC aims to find a single (optimal) trajectory,
CIPS simulate multiple stochastic trajectories in a Monte Carlo
manner.  Notably, the solution of the deterministic optimal control
problem~\eqref{eq:nonlinear_opt_control_problem} is based on
simulating~\eqref{eq:CIPS_intro} which is an SDE.  For the stochastic
MPC problems, multiple simulations have been considered in the
scenario-based approach ~\citep{campi2018introduction}.

\paragraph{Some perspectives on future research} In basic sciences,
there are a number of important examples of interacting particle
systems.  This paper presents results on the theme of 
``CIPS as an algorithm''.  The most 
historical of such algorithms is the EnKF which is used to solve the
problem of data assimilation.  It is hoped that this survey convinces
the reader that the paradigm is also useful for solving other problems in estimation
and control.  A
major selling point of CIPS, and also the reason for widespread use of
the EnKF, is that it is able to work directly with
a simulator.  Therefore, it is amenable as a solution method for
complex systems where models typically exist only in the form of a
simulator.  Apart from the open problems described in the main body of
the paper, a few themes for future research are as follows:
\begin{itemize}
\item MPC offers a useful benchmark for CIPS.  With the exception of
  the geometric approaches, e.g., FPF on Riemannian
  manifolds~\citep{zhang2017feedback}, constraints has not been an
  important theme in design of CIPS.  It is an important problem to
  extend the design of mean-field process to handle general types of
  constraints in inputs and states.  One possible next step is to
  extend the dual EnKF to the inequality-constrained LQR problems.  
\item RL could be an important application for CIPS.  A key difference
  is that CIPS-based solution does not rely on function
  approximation. Instead, the value function is approximated in terms
  of the distribution of the particles. This has some advantages,
  e.g., avoids the need to select basis functions, and some
  disadvantages, e.g., availability of computational resources.  It
  will be useful to understand some of these trade-offs.  
\item Relationship to mean-field games and optimal control should be
  further developed.  CIPS
  represent simple examples of mean-field type control laws.
  However, derivation of these control laws is, more often than not,
  rooted in methods from
  optimal transportation theory (\Sec{sec:OT-FPF}).  It remains an
  open problem to derive the FPF control law starting from a
  mean-field optimal control type objective (some partial results in this direction
  appear in~\citep{zhang2019mean}).   
\item {The lack of progress to obtain FPF as a solution of an
    optimal control problem is symptomatic of a satisfactory duality
    theory between optimal filtering and optimal
  control~\citep{todorov2008general}. Recent progress in
  this direction has been made in some work originating in our
  group~\citep{kim2019duality,JinPhDthesis,duality_jrnl_paper_II}.
  While the focus of this new work has thus far been on dual characterization
  of stochastic
  observability~\citep{duality_jrnl_paper_I,kim2019observability} and
  its use in filter stability
  analysis~\citep{kim2021ergodic,kim2021detectable}, it will be 
  interesting to explore connections both to FPF and to mean-field
  control. Duality-based derivation of the EnKF has previously been
  considered in~\cite{kim2018derivation}.}     
   \item Extensions to partially observed optimal control problems.
     For the linear Gaussian model, algorithms described in parts I
     and II are easily combined to obtain a CIPS for the partially
     observed problem.  The solution is based on the separation
     principle: A forward (in time) EnKF is run to solve the optimal filtering
     problem; and a completely independent backward (in time) dual
     EnKF is run to solve the optimal control problem.  For the
     nonlinear problem, there may be benefit to couple the forward and
     backward CIPS.    
  \item Distributionally robust FPF. In order to handle uncertainty in signal
    and observation models, it may be useful to explore methods from
    distributionally robust
    optimization framework~\citep{rahimian2019distributionally}.  The framework
    has been used to develop the Wasserstein robust Kalman filter for
    the linear Gaussian model~\citep{shafieezadeh2018wasserstein}. Its
    extension to the nonlinear filtering model~\eqref{eq:intro-model}
    is open and 
    may be possible based on the optimal transport formulation of
    the FPF.

\end{itemize}

	\bibliographystyle{elsarticle-harv} 
	\bibliography{references,add_refs,fpf_applns}

\end{document}